\documentclass[fleqn,3p,sort&compress]{elsarticle}

\usepackage{amsmath}
\usepackage[breaklinks]{hyperref}
\usepackage[latin1]{inputenc}
\usepackage{graphicx}
\usepackage{color}
\usepackage{amsfonts,amsthm,amssymb}
\usepackage{bm,bbm}
\usepackage[OT2,OT1]{fontenc}
\usepackage{slashed}
\usepackage{verbatim}
\usepackage[usenames,dvipsnames]{xcolor}
\usepackage{axodraw}

\newcommand\cyr{%
\renewcommand\rmdefault{wncyr}%
\renewcommand\sfdefault{wncyss}%
\renewcommand\encodingdefault{OT2}%
\normalfont
\selectfont}
\DeclareTextFontCommand{\textcyr}{\cyr}

\def\beq{\begin{equation}}
\def\eeq{\end{equation}}
\newcommand{\be}{\begin{eqnarray}}
\newcommand{\ee}{\end{eqnarray}}

\renewcommand{\texttt}{{}}

\def\bs{\begin{subequations}}
\def\es{\end{subequations}}

\newcommand{\tia}[1]{}

\begin{document}

\begin{frontmatter}
\title{
{\bf{%Four gravitons 
%Tree level Graviton s
Scattering amplitudes in super-renormalizable gravity
%local \& nonlocal higher derivative gravity
}}}

\author{Pietro Don\`a} 
\ead{pietro$\_$dona@fudan.edu.cn}

\author{Stefano Giaccari} 
\ead{giaccari@fudan.edu.cn}

\author{Leonardo Modesto} 
\ead{lmodesto@fudan.edu.cn}

\author{Les\l aw Rachwa\l{}}
\ead{rachwal@fudan.edu.cn}

\author{Yiwei Zhu}
\ead{12210190003@fudan.edu.cn \quad\quad\quad\quad}

\address{
{\small Department of Physics \& Center for Field Theory and Particle Physics,} \\
{\small Fudan University, 200433 Shanghai, China}
}

\date{\small\today}

\begin{abstract} \noindent
We explicitly compute the tree-level on-shell four-graviton amplitudes in four, five and six dimensions 
for local and weakly nonlocal gravitational theories that are quadratic in both, the Ricci and scalar curvature
with form factors of the d'Alembertian operator inserted between.  
More specifically we are interested in renormalizable, super-renormalizable or finite theories. 
The scattering amplitudes for these theories turn out to be the same as the ones of Einstein gravity regardless of the explicit form of the form factors. As a special case the four-graviton scattering amplitudes in Weyl conformal gravity are identically zero. 
Using a field redefinition, we prove that the outcome is correct for any number of external gravitons (on-shell $n-$point functions) and in any dimension for a large class of theories. However, when an operator quadratic in the Riemann tensor is added in any dimension (with the exception 
of the Gauss-Bonnet term in four dimensions) the result is completely altered, and the scattering amplitudes depend on all the form factors introduced in the action. 
\end{abstract}

\end{frontmatter}

\tableofcontents

\section{Introduction}
Scattering amplitudes, in particular gauge invariant on-shell ones, are the observables of major interest in particle physics experiments. 
The possibility of getting very precise and well testable predictions for them is surely one of the biggest achievements of perturbative quantum field theory. From a more theoretical point of view, they play a fundamental role in constraining the higher derivative terms that are expected to arise at quantum level as a consequence of the presence of divergences in loop diagrams in non-renormalizable theories \cite{Wang:2015jna}.
It is well known that, even for tree-level diagrams, the intermediate steps of computation show a remarkable 
tangle that often disappears in the final physical amplitudes \cite{DeWitt}. This is a consequence of the ambiguity (and also the richness) intrinsic to the Lagrangian off-shell formalism, which is invariant under gauge transformations and field redefinitions.  Recently there has been a wide interest in developing powerful methods for on-shell scattering amplitudes, which have led to both improving calculation techniques and gaining new insights into the underlying mathematical structures (see \cite{Elvang:2013cua} and references therein). 

On the other hand, since it is quite difficult to propose good and unambiguous quantities which could play the role of observables in pure quantum gravity, we concentrate our attention on hypothetical experiments of graviton scattering described in the perturbative framework of quantum field theory around flat Minkowski spacetime. Due to the kinematical constraints the first non-vanishing amplitude (for on-shell massless gravitons) must describe a scattering process involving four particles. One example is the scattering of two gravitons into two, which is the topic of this paper. Here due to tremendous complications present at loop levels in quantum gravity, we consider only tree-level amplitudes (only in the Born approximation). Physically these amplitudes correspond to the classical processes of monochromatic gravitational wave scattering happening in the empty space. Of course such effects are predicted by Einstein general relativity, which is a nonlinear theory of gravitational interactions, but non-linearities are very little effects due to the smallness of the ratio 
between the typical graviton energy and the Planck mass. Nonetheless, the amplitudes are indeed very useful probes of quantum gravitational physics and are the first must-be-taken step to quantum gravity phenomenology.

The first attempt to compute tree-level scattering amplitudes was undertaken in the simplest quantum gravity theory, namely in the quantum version of Einstein gravity. There it was found \cite{DeWitt, EHcomputations} 
that the final results are quite simple and can be derived even without performing detailed computations using Feynman diagram methods \cite{Grisaru:1975bx}. Due to dimensional reasons ($G_N$ has energy dimension $-2$ in four dimensions) the amplitudes showed behavior that grow unboundedly with increasing energy of the scattering process. Precisely they grow as $E^2$. This observation led to the conclusion 
that naive tree-level unitarity bound on the scattering amplitudes is violated. Further, it was derived that either Einstein-Hilbert quantum gravity becomes non-perturbative around Planck scale or the theory needs an ultraviolet (UV) completion and can be viewed merely as a low energy effective field theory of quantum gravity. We keep this message in mind and we study the scattering processes in a class of theories which can be without problems in the UV regime, but at the same time these theories remain always in the perturbative regime. 
Indeed, higher derivative super-renormalizable theories are asymptotically free and the unitarity bound is 
not violated. This is transparent in the prototype asymptotically free theory proposed by Stelle in 1977
\cite{Stelle, TomboAF, Fradkin}. 
The energy scaling of the scattering amplitudes is proportional to $E^4$ or $E^2$, but at a closer inspection it can be correct only up to the Planck scale because 
at very high energy the interaction between gravitons becomes very weak and they travel almost as free particles
as a mere consequence of asymptotic freedom. 
Therefore, the scattering amplitudes presented in this paper are meaningful up to the Planck scale
(or the new scale introduced in the theory), 
while a careful analysis is required around the Planck mass.

Moreover, despite the remarkable result of our computation, which states that amplitudes in four dimensions are the same as in Einstein gravity, we could be able to determine other phenomenological consequences of the graviton scattering processes. One of such interesting implications concerns the production of micro black holes as resonances in the elastic scattering of gravitons \cite{Calmet1, Calmet2}. Typically, as it is done in high energy physics, from a broadening of the cross section we would be able to read out the half-width and therefore determine the life-time of such objects. This would shed a new light onto the interpretation of Hawking evaporation process in the fully-fledged (opposed to semi-classical approach) quantum field theory of gravitational interactions. Other implications for particle physics and its unification with gravity would require the coupling of matter to gravity, which we do not attempt to make here.

As the last piece of our original motivation we want to mention that the results of the amplitude computation might have served well for the determination of form factors, which are used in nonlocal theories (for the special case of constant form factors we would be able to determine values of the couplings in front of terms quadratic in curvatures in the Stelle gravity). But since in four spacetime dimensions we find no dependence on interpolating functions (form factors) with the exception of terms quadratic in Riemann tensors, this cannot be done based only on tree-level results for the minimal super-renormalizable or finite theory. The situation is different in higher dimensions as we point out explicitly in the second part of the paper and this is one of the new results of this paper. In the higher dimensional setup, thanks to these amplitudes, we are able to determine some characteristic features of the form factors. For fully unambiguous determination of the form factors in the minimal nonlocal super-renormalizable theory we will need the full one-loop result, which is right now beyond our capabilities. It can be quite pleasing that in four physical dimensions tree-level amplitudes do not depend on the form factors, because then we can view the presence of the form factors as the clever way of parametrizing our ignorance. However, as we see on the tree-level nothing depends on it and actually we are not forced to determine the form factor there. This changes one of the biggest drawback of the theory (ambiguity related to the choice of the form factor) into a virtue (now tree-level scattering amplitudes are independent of it). This further may lead to seeing the form factors as something that does not have very important physical meaning. The form factor may be used only as a way to interpolate between infrared (IR) and UV behavior of the theory in a covariant way. Such ideas are similar to the ideas of the smooth covariant cutoff developed in the field of functional RG. We believe that the true significance of the form factors used in nonlocal theories will be revealed only at the loop level. This is confirmed by partial computations of the beta functions of running couplings, which indeed do depend on the asymptotic form of the form factors used.

All of this has drawn our attention to investigations, with somewhat more traditional techniques, of what kind of physical information can be obtained from tree-level on-shell amplitudes for a class of weakly nonlocal theories of gravity, which have been the subject of recent studies \cite{modesto, modestoLeslaw, universality, Briscese, Modesto:2013jea, Piva, Eran, Krasnikov, Kuzmin, Tombo}. These theories turn out to be a particularly convenient theoretical framework to perform quantum gravity computations, because they have been proven to be ghost-free and super-renormalizable or finite at quantum level.
We work in the quantum field theory framework and we assume as our guiding principle the 
``validity of perturbative expansion" \cite{ANSELMI}. 
Moreover, the following postulates are required:
(i) spacetime diffeomorphism covariance; (ii) weak nonlocality (or quasi-polynomiality); (iii) unitarity (the spectrum is tachyon- and ghost-free); (iv) super-renormalizability or finiteness.  
The main difference with perturbative quantum Einstein gravity lies in the second requirement, which makes possible to achieve 
unitarity and renormalizability at the same time. 
In this way a fully consistent quantum gravitational theory has been constructed 
free of any divergences. However, the theory is not unique and all our ignorance is encoded 
in a form factor (entire function) with very specific asymptotic limits in the ultraviolet and in the infrared fixed in such a way to have a convergent quantum field theory. At classical level there are evidences that we are dealing with a 
``{\em singularity-free}" gravitational theory in the case of physical matter
 \cite{Moffat3, corni1, ModestoMoffatNico,BambiMalaModesto2, BambiMalaModesto,calcagnimodesto, koshe1, Modesto:2014eta}. 
However, we know that Einstein spaces are exact vacuum solutions of the weakly nonlocal theory, including the singular Schwarzschild spacetime \cite{YaoDong, AnselmiCS}. 

The main findings of this paper consist of new results for tree-level scattering amplitudes in Stelle gravity and its analytic nonlocal extensions. We find that the situation in quadratic gravity in four dimensions is exactly the same as in Einstein gravity, while for the Weyl conformal theory the four-graviton amplitudes are identically zero. In dimension higher than four and when the terms quadratic in Riemann tensors are included the outcome in Stelle gravity differs from the standard Einstein theory. If in four dimensions (or extra dimensions) we allow for analytic nonlocal extensions involving the Riemann tensor, then again we find a different outcome. We finally explain our results by proving a general theorem about on-shell $n-$graviton scattering amplitudes and by explicitly considering diagrams from terms that cannot be redefined. We point out the possibility of determining the form factors, when these theories are viewed as fundamental, and comparing them with the experimental results about graviton scattering amplitudes. This may be taken as a first step to verify this class of theories.

In this paper we mainly concentrate on the calculation of tree-level
four-graviton scattering amplitudes. In the remaining part of this introduction we remind the reader the construction of the weakly nonlocal quantum gravity. Later, in the second section, we start with some kinematical considerations and we find the propagator and compact expressions for vertices. We first perform the amplitude computation in quadratic Stelle theory \cite{Stelle} in section three and then in the most general weakly nonlocal super-renormalizable theory quadratic in the Ricci and scalar curvature tensors in section four. Moreover, in the fifth section we interpret the technical results as a consequence of a general theorem
based on Anselmi's field redefinition theorem \cite{Anselmi:2006yh, Anselmi:2002ge}, and 
\cite{Marcus:1984ei, 'tHooft:1974bx, Goroff:1985th, Deser:1986xr}. Finally, we discuss the generalizations of our results (also some expectations beyond tree-level) and we give conclusions. We supplement the paper by two appendices, where we put more technical details about the propagator and variations of curvature invariants.

\subsection{Weakly nonlocal gravity}
The general $D$-dimensional theory weakly nonlocal and quadratic in Riemann, Ricci and scalar
curvature reads 
\cite{modesto,modestoLeslaw, Briscese, Krasnikov, Tombo, BM, M3, M4, Khoury:2006fg, Cnl1, Mtheory,NLsugra, Modesto:2013jea},
\be
&& 
\mathcal{L}_{\rm g} = -  2 \kappa_{D}^{-2} \, \sqrt{-g} 
\left[ {\bf R }
+
{\bf R }\, 
 \gamma_0(\square)
{\bf  R }
 + {\bf Ric} \, 
\gamma_2(\square)
 {\bf Ric} 
+ {\bf Riem}  \, 
\gamma_4(\square) 
{\bf  Riem} 
+ {\bf V} \, 
\right]  \,.
\label{gravity}
\ee
The theories above consist of a weakly nonlocal kinetic operator and a local curvature potential ${\bf V}$. Therefore they are quite general, more general thing would be only to allow the potential to be nonlocal too. The local potential ${\bf V}$ for a gravitational theory, which is however sufficient, is made up of the following three sets of operators 
\be
\hspace{-0.6cm}
 \sum_{j=3}^{{\rm N}+2} \sum_{k=3}^{j} \sum_i c_{k,i}^{(j)} \left( \nabla^{2(j-k)} {\cal R}^k \right)_i
 +
  \!\!\!
 \sum_{j={\rm N}+3}^{\gamma+{\rm N}+1} \sum_{k=3}^{j} \sum_i d_{k,i}^{(j)} \left(\nabla^{2(j-k)} {\cal R}^k \right)_i
 + \!\!\!
 \sum_{k=3}^{\gamma +{\rm N}+2} \sum_i s_{k,i} \, \left(  \nabla^{2 (\gamma + {\rm N}+2 -k )} \, {\cal R}^k \right)_i ,  
 \label{K0}
 \ee

\noindent where operators in the third set are called killers, because they are crucial in making the theory finite in any 
 dimension. They are used to kill the beta functions.
 In (\ref{K0}) Lorentz indices and tensorial structure have been neglected\footnote{{\em Definitions ---} 
The metric tensor $g_{\mu \nu}$ has 
signature $(- + \dots +)$ and the curvature tensors are defined as follows: 
$R^{\mu}_{\nu \rho \sigma} = - \partial_{\sigma} \Gamma^{\mu}_{\nu \rho} + \dots $, 
$R_{\mu \nu} = R^{\rho}_{\mu  \rho \nu}$,  
$R = g^{\mu \nu} R_{\mu \nu}$. With symbol ${\cal R}$ we generally denote one of the above curvature tensors.}. 
For more details about notation in the potential ${\bf V}$ we refer the interested reader to original papers \cite{modestoLeslaw, universality}. Moreover  in (\ref{gravity})
$\square = g^{\mu\nu} \nabla_{\mu} \nabla_{\nu}$ is the covariant box operator, and the functions 
$\gamma_\ell(\square)$ ($\ell=0,2,4$) (form factors) will be shortly defined. 
When $\gamma_4(\square)=0$ minimal unitarity requires 
\be
\gamma_0(\square) = - \frac{\gamma_2(\square)}{2} =  
- \frac{e^{H(-\square_{\Lambda})} -1}{2 \, \square}\,,
\label{unitarity1}
\ee
where $H(z)$ is an entire function on the complex plane.
Given the conditions (\ref{unitarity1}), in the perturbative spectrum of the theory we only have the massless transverse graviton, whose propagator is modified only in a multiplicative way from the Einstein theory.
In (\ref{unitarity1}) $\Lambda$ is a fundamental invariant mass scale in our theory and we define $z=-\square_{\Lambda}=-\square/\Lambda^2$.

A universal exponential form factor $\exp H(z)$ compatible with the guiding principles of quantum field theory and with the requirement of proper IR limit of the theory $H(0)=0$ is \cite{Tombo}:
\be
&&V^{-1}(z)=e^{H(z)}
= e^{\frac{a}{2} \left[ \Gamma \left(0, p(z)^2 \right)+\gamma_E  + \log \left( p(z)^2 \right) \right] } \nonumber\\
&&
=
e^{a \frac{\gamma_E}{2}} \,
\sqrt{ p(z)^{2 a}} 
\left\{ 
1+ \left[ \frac{a \, e^{-p(z)^2}}{2 \,  p(z)^2} \left(  1  
+ O \left(   \frac{1}{p(z)^2} \! \right)   \! \right) + O \left(e^{-2 p(z)^2} \right)  \right] \right\}  , 
 \label{Tomboulis}
\ee
where the last equality is correct only on the real axis. 
 $\gamma_E \approx 0.577216$ is the Euler-Mascheroni constant and 
$
\Gamma(0,z) = \int_z^{+ \infty}  d t \, e^{-t} /t 
$ 
is the incomplete gamma function with its first argument vanishing. Now we define capital~$\rm{N}$ by the following function of the spacetime dimension $D$: $2 \mathrm{N} + 4 = D$ in even and $2 \mathrm{N} + 4 = D+1$ in odd dimensions. Besides integer $\rm{N}$ in our theory we have also another integer number $\gamma$, which measures how far the theory is from the minimal renormalizable one in given spacetime dimension.
The  polynomial $p(z)$ in (\ref{Tomboulis}) is of degree $\gamma +\mathrm{N}+1$ and such that $p(0)=0$, which gives the correct low energy limit of our theory (coinciding with Einstein gravity).
The entire function has asymptotic UV behavior given by the polynomial $p(z)^a$ in conical regions around the real axis. However in most applications we will choose a monomial behavior $z^{a(\gamma + {\rm N} +1)}$, where $a$ is another positive integer number. Then these conical regions are with an opening angle
$\Theta= \pi/(4 (\gamma + \mathrm{N} + 1))$. For $\gamma ={\rm N}=0$ we have the maximal angle $\Theta = \pi/4$ for all $a$.

\subsection{Propagator, tree-level unitarity and power counting} \label{gravitonpropagator}
We shortly discuss the propagator in a theory (\ref{gravity}). Splitting the spacetime metric into the Minkowski background plus fluctuations and after performing a gauge fixing, we can invert the quadratic in fluctuation fields kinetic operator to finally get the two-point function  in Fourier space \cite{HigherDG}
(see section (\ref{propagator2}) and \ref{appendixProp}  for more details),
\be
&& \hspace{-1cm}   \mathcal{O}^{-1} \!=\!
 \frac{V(  k^2/\Lambda^2 )  } {k^2} \left( P^{2} - \frac{P^{0\theta}}{D-2 }  \right) .
   \label{propagator}
\ee
The indices for the operator $\mathcal{O}^{-1}$ and the projectors 
\cite{HigherDG, VN} $\{ P^{0\theta},P^{2}
\}$ as well as gauge-dependent terms in (\ref{propagator}) have been omitted. 
The tensorial structure in (\ref{propagator}) is the same of Einstein gravity, but the multiplicative 
 form factor $V(  k^2/\Lambda^2 )$ in Fourier space makes the theory strongly ultraviolet (UV) convergent without the need to modify the spectrum or introducing ghost instabilities. The theory propagates only massless spin two particle (graviton with two helicities), however the UV behavior of the propagator is modified. This fact about the tree-level spectrum ensures us that we are dealing with a perturbatively unitary theory.

We now review the power counting analysis of the quantum divergences \cite{Kuzmin, Tombo, Eran, modesto}. 
In the high energy regime, and in even dimension
the above propagator (\ref{propagator}) in momentum space 
schematically scales as 
\be
\mathcal{O}^{-1}(k) \sim \frac{1}{k^{2 \gamma +D}} \,\,\,\,\,\,
\mbox{in the UV} \, . 
\label{OV} 
\ee
The interaction vertices can be collected in different sets, that may or may not involve the entire functions $\exp H(z)$. However, to find a bound on the quantum divergences it is sufficient to concentrate on
the leading operators in the UV regime. 
These operators scale as the propagator giving the following 
upper bounds on the superficial degree of divergence of any graph \cite{modesto,A}, 
\be
\omega(G)=DL+(V-I)(2 \gamma + D)\,.
\ee 
We rewrite above in a more convenient form as
\be 
\omega(G) =  D - 2 \gamma  (L - 1)    \, . \,\,\,\,
\label{even}
\ee
In 
(\ref{even}),  we used the topological relation between the numbers of vertices $V$, internal lines $I$ and the
number of loops $L$: $I = V + L -1$. 
Thus, if $\gamma > D/2$, in the theory only 1-loop divergences survive.  
Therefore, 
the theory is super-renormalizable \cite{Krasnikov, Efimov, Moffat3,corni1}
and only a finite number of operators of mass dimension up to $M^D$ has to be
included in the action in even dimension. For odd dimension, if $\gamma > (D-1)/2$, then the theory is completely without divergences and hence automatically finite.

 \subsection{Super-renormalizable \& finite gravitational theories in $D=4$}\label{FINITA}
The main reason to introduce a potential $\bf V$ in the action (\ref{gravity}) is to make the theory finite at quantum level.
It is easy to see that it is always possible to choose the non-running coefficients $s_{k,i}$ in (\ref{K0})
to make all the beta functions vanish. We consider the simplest case of a 
monomial asymptotic behavior for the form factor, namely: $p_{\gamma + {\rm N} +1} (z) = z^{\gamma + {\rm N} +1}$. For this particular choice of the form factor in the large $z$ limit, the analysis of the previous subsection shows that only divergences, which are to be renormalized by terms with $D$ derivatives 
(like $\mathcal{R}^{D/2}/\epsilon$ in dimensional regularization (DIMREG)), are generated at one-loop.
However, the killer operators in the last set of operators in (\ref{K0}) give  contributions to the beta functions of the theory linear in the front parameters $s_{k,i}$. The latter ones can be fixed in such a way to make the theory finite in any even dimension. The tensorial structure of killers must reflect the structure of terms, which are renormalized in the original theory (terms with $D$ derivatives).

In $D=4$ (${\rm N}=0$) the whole situation is simple 
 to describe, because we only need two killer operators. 
The highest derivative terms in the kinetic part of
the action come from the form factor and 
are of the type ${\cal R} \, \square^{\gamma}{\cal R}$ \cite{Shapirobook, shapiro3, HigherDG}. 
The minimal choice for a finite and unitary theory of quantum gravity in four dimensions
may consist of terms with $\gamma=3$ (and $a=1$) in the kinetic part. This alone leads to one-loop super-renormalizable quantum nonlocal gravity. For simplicity we introduce only
two quartic killers and no cubic in curvature operators. This is sufficient to make vanish all two beta functions for $R^2$
and $R_{\mu\nu}^2$ operators. The simplest Lagrangian with the Tomboulis type of form factor (\ref{Tomboulis}) 
may be the following,
\be
 \mathcal{L}_{\rm fin} = - \frac{2}{ \kappa_4^{2}} \sqrt{ - g}   \Big[ R  
+ R_{\mu \nu} \,  \frac{ e^{H(-\square_{\Lambda})} -1}{\square}   R^{\mu \nu} 
-  R \,  \frac{ e^{H(-\square_{\Lambda})} -1}{2 \square}   R
+s_{1}R^{2}\square (R^{2})+s_{2}R_{\mu\nu}R^{\mu\nu}\square (R_{\rho\sigma}R^{\rho\sigma}) \Big]  ,
\label{Minimal}
\ee
where $p(z) =z^4$,
$s_{1}=-2\pi^2\omega_{2}(c_{1}+c_{2})/3$, $s_{2}=8\pi^2\omega_{2}c_{2}$ and 
$\omega_2 = \Lambda^{-8} \exp ({\gamma_E/2})$ \cite{modestoLeslaw}.
Here $c_{1}$ and $c_{2}$ are two constants independent on $\omega_{2}$,
that have to be determined from the calculation of the contributions to the
beta functions from terms quadratic in curvature and dominant at high energies. A more general Lagrangian can have a finite number of other local terms, but still finiteness of the theory can be obtained exactly in the same way.

We recently proposed, following \cite{Kuzmin}, another class of weakly nonlocal possibly finite theories, which are constructed entirely from kinetic terms (only 
weakly nonlocal operators quadratic in curvature appear), without local or nonlocal gravitational potential 
${\bf V}$ cubic in curvature or higher. 
The simplest nonlocal four-dimensional theory we can write to achieve finiteness is exactly the one given 
in (\ref{gravity}), but without assuming the relation (\ref{unitarity1}) 
(unitarity is anyhow achieved as proven in \cite{Briscese, Piva}), 
\be
&&
\mathcal{L} = - 2 \kappa^{-2}_4 
\sqrt{ - g} [  R + R \gamma_0(\square) R + R_{\mu\nu} \gamma_2(\square) R^{\mu\nu}
+ R_{\mu\nu\rho\sigma}  \gamma_4(\square) R^{\mu\nu\rho\sigma} 
 \, ] \, ,  \quad\label{elegantfin}  \\
&& \gamma_\ell(\square) = \frac{e^{H_\ell(-\square_\Lambda)} - 1}{\square} \, \nonumber. 
\ee
We assume that all three form factors (for $\ell=0,2$ and 4) have the Tomboulis form (\ref{Tomboulis}) with the same degree of UV polynomial $\gamma+1$. The operators in (\ref{elegantfin}) can be written equivalently in other bases using Weyl tensors $C_{\mu\nu\rho\sigma}  \gamma_w(\square) C^{\mu\nu\rho\sigma}$ or generalized Gauss-Bonnet Lagrangian. The finite theory would boil down to some relations between UV polynomials for each form factor. At the moment we cannot definitely assert that the theory is finite, because the beta functions for the operators $R^2$
and $R_{\mu\nu}^2$ are quadratic in three parameters defining the asymptotic UV
behaviors of the entire functions $H_\ell$. 
This theory and its generalizations 
are currently under careful investigation
and the results will be published in a future paper.
Nevertheless, looking at the beta functions reported by Kuz'min \cite{Kuzmin}, it turns out that for $\gamma>4$ 
we can always  find a solution for 
\be
\beta_{R^2}=0 \,\,\, {\rm or} \,\,\, \beta_{R_{\mu\nu}^2}=0 \, , 
\ee
hence in this class of theories only one beta function could be non-zero.

\section{Four-graviton scattering amplitudes in higher derivative gravity} %Scattering 
We start by studying  the higher derivative gravity theory
proposed and extensively studied by Stelle in 1977. 
The action is quadratic in curvature and it is the limiting case of (\ref{gravity}) for the particular case of constant form factors and zero potential, namely: 
\be
\gamma_0(\square) = {\rm const} = \gamma_0\, , \quad \gamma_2(\square) = {\rm const} = \gamma_2 \, , 
 \quad \gamma_4(\square) = {\rm const} = \gamma_4\, ,  \quad {\bf V} =0 \, .
\ee 
Therefore, the higher derivative Stelle's gravitational action reads
\be
&& 
S_{\rm g} = -  2 \kappa_{D}^{-2} \,\int \! d^D x \, \sqrt{-g} 
\left[ R 
+ \gamma_0 \, 
 R^2 
 + 
\gamma_2
R_{\mu\nu}^2 
+ 
\gamma_4 R_{\mu\nu\rho\sigma}^2 \right]  .
\label{gravityStelle}
\ee
In this section, before carrying out the computation of the on-shell four-graviton scattering amplitudes, we derive the propagator and vertices and discuss some  general properties of helicity amplitudes.

\subsection{Helicity amplitudes}
\label{helicityamp}

For the external on-shell gravitons we assume the physical conditions in transverse traceless gauge
\be
\partial^{\mu}h_{\mu\nu}=h_{\mu}^{\mu}=0 \, ,
\label{onshellgrav}
\ee
 which will turn
out to be very convenient in order to simplify the algebra all along
the computation. In this gauge \cite{EHcomputations}   the polarization tensors for gravitons with helicities
$\pm2$ in four dimensions are related to those of photons with helicities $\pm1$ by tensor product
\be
\label{poltensor}
\epsilon_{\mu\nu}(p,\pm2)=\epsilon_{\mu}(p,\pm1)\epsilon_{\nu}(p,\pm1) \, ,
\ee
where the polarization vectors satisfy
\begin{eqnarray}
 &&  \epsilon_{\mu}(p,\lambda)p^{\mu}=0,\quad \epsilon_{\mu}(p,\lambda)\epsilon^{\mu}(p,\lambda)=0 \, , \quad
 \nonumber \\
 &&
  \epsilon_{\mu}(p,-\lambda)=\epsilon_{\mu}^{*}(p,\lambda),
  \quad \epsilon_{\mu}(p,\lambda)\epsilon^{\mu}(p,-\lambda)=1 \, . 
  \label{eq:VectPolConditions}
\end{eqnarray}
Furthermore, we require the polarization vectors to form a complete basis
for a representation of the $SO(2)$ group of rotations in the transverse
directions which leave $p^{\mu}$ invariant. 
In the axial gauge we introduce the auxiliary vector $q^{\mu} \nparallel p^{\mu}$ such
that $q\cdot p\neq0$, and the sum on the polarizations reads
\begin{equation}
\sum_{\lambda=\pm 1}\epsilon_{\mu}(p,\lambda)\epsilon_{\nu}^{*}(p,\lambda)=\eta_{\mu\nu}-\frac{p_{\mu}q_{\nu}+p_{\nu}q_{\mu}}{p\cdot q} \, . 
\label{eq:CompletenessRel}
\end{equation}
In particular, taking momenta in spherical coordinates $(\bar\theta,\,\phi)$ to be
\be 
\label{prototypemomentum}
p_{\mu}=\left(p_{0},p_{0}\sin\bar\theta\cos\phi,p_{0}\sin\bar\theta\sin\phi,p_{0}\cos\bar\theta\right),
\ee
we can make the following explicit choice of the polarization basis, 
\be
\label{prototypepolarization}
\epsilon_{\mu}(p,\pm 1)=\frac1{\sqrt{2}}\left(0,\cos\bar\theta\cos\phi\mp i\sin\phi,\cos\bar\theta\sin\phi\pm i\cos\phi,-\sin\bar\theta\right)
\, .
\ee
Therefore we choose the polarization vectors to be helicity eigenvectors.
For the special choice $q_\mu=\left(p_{0},-\vec{p}\right)$ they satisfy
(\ref{eq:VectPolConditions}) and (\ref{eq:CompletenessRel}). 

We denote by $F_{\lambda_{3},\lambda_{4};\lambda_{1},\lambda_{2}}$
the helicity amplitudes for the scattering process $1+2\rightarrow3+4$
of spin $2$ particles with helicities $\lambda_{i}$. The helicity amplitudes
satisfy a number of relations as a consequence of parity conservation
and time reversal for a reaction $a+b\rightarrow a+b$, Bose symmetry
for a reaction $a+a\rightarrow b+c$, and invariance under particle-antiparticle
conjugation for a reaction $a+\bar{a}\rightarrow b+\bar{b}$, namely
\begin{eqnarray}
&& F_{\lambda_{3},\lambda_{4};\lambda_{1},\lambda_{2}}  =  \left(-1\right)^{\lambda-\mu}F_{-\lambda_{3},-\lambda_{4};-\lambda_{1},-\lambda_{2}} \, , \nonumber \\
&& F_{\lambda_{3},\lambda_{4};\lambda_{1},\lambda_{2}} = \left(-1\right)^{\lambda-\mu}F_{\lambda_{1},\lambda_{2};\lambda_{3},\lambda_{4}} \, ,
\nonumber \\
&& F_{\lambda_{3},\lambda_{4};\lambda_{1},\lambda_{2}} =  \left(-1\right)^{\lambda-4} 
 F_{\lambda_{3},\lambda_{4};\lambda_{2},\lambda_{1}}  \,  ,
 \nonumber \\
&& F_{\lambda_{3},\lambda_{4};\lambda_{1},\lambda_{2}}  =  \left(-1\right)^{\lambda-\mu}F_{\lambda_{3},\lambda_{4};\lambda_{2},\lambda_{1}}\, , 
\label{eq:HelicityRel}
\end{eqnarray}
where $\lambda=\lambda_{1}-\lambda_{2}$ and $\mu=\lambda_{3}-\lambda_{4}$.

The most significant advantage of working with such amplitudes is
that they are defined as Lorentz invariant quantities for massless particles,
which implies we are automatically computing gauge invariant quantities,
only dependent on the Mandelstam invariants $s=-\left(p_{1}+p_{2}\right)^{2}$, $t=-\left(p_{1}-p_{3}\right)^{2}$ \\
and $u=-\left(p_{1}-p_{4}\right)^{2}$.

Using relationships  \eqref{eq:HelicityRel}, out of 16 amplitudes in four dimensions, we can single out four independent helicity amplitudes \cite{Grisaru:1975bx} 
\begin{equation}
\begin{array}{lll}
\hspace{2cm}&\mathcal{A}\left(++,++\right)\equiv F_{2,2;2,2} , &  \mathcal{A}\left(+-,+-\right) \equiv F_{2,-2;2,-2}, \\
&\mathcal{A}\left(++,+-\right) \equiv F_{2,2;2,-2} , & \mathcal{A}\left(++,--\right) \equiv F_{2,2;-2,-2}.
\label{helicity}
\end{array}
\end{equation}

\subsection{The two-point function} \label{propagator2} 
Let us start by considering the propagator of the Einstein-Hilbert
theory assuming the expansion of the metric around a flat background
\begin{equation}
g_{\mu\nu}=\eta_{\mu\nu}+ h_{\mu\nu} \,  ,
\end{equation}
where $h_{\mu\nu}$ is the fluctuation field. We will denote by $\mathcal{O}^{(n)}$ the $n$-th order term of the expansion of operator $\mathcal{O}$ in the fluctuation field. Furthermore we will distinguish between on-shell and off-shell fields by underlining the latter.

The equation of motion is given by
\be
{S}_{\rm EH}=-2\kappa_{D}^{-2} \int\! d^{D}x\,\sqrt{-g}R \quad \Longrightarrow  \quad 
\frac{\delta {S}_{\rm EH}}{\delta g_{\mu\nu}} \delta g_{\mu\nu} = - 2\kappa_{D}^{-2} \int\! d^{D}x\, \sqrt{-g}\left(\frac{1}{2}g^{\mu\nu}R-R^{\mu\nu}\right) \delta g_{\mu\nu}\, .
\ee
We can always assume $(\delta{S}_{\rm EH}/\delta g_{\mu\nu})^{(0)}=0$
because $R^{(0)}=R_{\mu\nu}^{(0)}=R_{\mu\nu\rho\sigma}^{(0)}=0$ as a consequence of having chosen a flat background metric. 
The two point function is obtained by considering the expansion to the second order 
\be
\hspace{-0.2cm}
{S}^{(2)}_{\rm EH}=-2\kappa_{D}^{-2} \int\! d^{D}x\left(\sqrt{-g}R\right)^{(2)} 
= -2\kappa_{D}^{-2} \int\! d^{D}x\left[\left(\sqrt{-g}\right)^{(2)}R^{(0)}+\left(\sqrt{-g}\right)^{(1)}R^{(1)}+\left(\sqrt{-g}\right)^{(0)}R^{(2)}\right]  .
\ee
%
\begin{comment}
\be \hspace{0.1cm}
\frac{1}{2} \underline{h}_{\alpha\beta}(x)\left(\frac{\delta^{2}{S}_{EH}}{\delta g_{\alpha\beta}(x)\delta g_{\gamma\delta}(y)}\right)^{\!\!(0)}  \!\!\!\!
 \underline{h}_{\gamma\delta}(y)
=  \frac{1}{2} \underline{h}_{\alpha\beta}(x)\left(-\frac{2}{\kappa_{D}^{2}}\right)
\left\{
\frac{\delta}{\delta g_{\alpha\beta}(x)}\left(\sqrt{-g}\left(\frac{1}{2}g^{\gamma\delta}R-R^{\gamma\delta}\right)\right)\right\}^{\!\!(0)} \!\!\!
 \underline{h}_{\gamma\delta}(y) \, ,
\ee
where we introduce the underline notation in referring at the off-shell gravitons. 
\end{comment}
For an on-shell graviton
\be
&& R^{(1)}  =  \partial_{a}\partial_{b}h^{ab}-\square h_{a}^{a} =0\, , \nonumber \\
&& R_{\mu\nu}^{\left(1\right)}  =  \frac{1}{2}\left(-\partial_{\mu}\partial_{\nu}{h}_{a}^{a}+\partial_{\mu}\partial^{a}{h}_{a\nu}+\partial_{\nu}\partial^{a}{h}_{a\mu}-\square{h}_{\mu\nu}\right) = 0\,, \nonumber \\
&&R_{\mu\nu\varrho\sigma}^{(1)} =  \frac{1}{2}\left(-\partial_{\mu}\partial_{\rho}h_{\nu\sigma}+\partial_{\mu}\partial_{\sigma}h_{\nu\rho}
 +\partial_{\nu}\partial_{\rho}h_{\mu\sigma}-\partial_{\nu}\partial_{\sigma}h_{\mu\rho}\right) \neq 0\,,
\ee
which in particular imply the minimum condition for the quadratized Einstein-Hilbert action 
\be
\left(\frac{\delta^{2}{S}_{\rm EH}}{\delta g_{\alpha\beta}(x)\delta g_{\gamma\delta}(y)}\right)^{\!\!(0)}=0\, .
\ee
%
%STELLE
For  the higher derivative theory (\ref{gravityStelle}) we need to add the second order expansion of the quadratic terms ${S}_{\rm QUAD}$
\be
{S}^{(2)}_{\rm QUAD}=\left(-2\kappa_{D}^{-2}\right) \int\! d^{D}x \left(\sqrt{-g}\right)^{(0)} \left(\gamma_{0}R^{(1)2}+\gamma_{2}R_{\mu\nu}^{(1)2}+\gamma_{4}R_{\mu\nu\rho\sigma}^{(1)2}\right) \, .
\ee

\begin{comment}
\be
&& \hspace{-1.2cm}
\frac{1}{2} \underline{h}_{\alpha\beta}(x)\left(\frac{\delta^{2}{S}_{g}}{\delta g_{\alpha\beta}(x)\delta g_{\gamma\delta}(y)}\right)^{\!\!(0)}
 \!\!\! \underline{h}_{\gamma\delta}(y) 
 =  \frac{1}{2}  \underline{h}_{\alpha\beta}(x) 
  \left(-\frac{2}{\kappa_{D}^{2}} \right)\left[ \frac{1}{2}\eta^{\gamma\delta}\left(\frac{\delta R(y)  }{\delta g_{\alpha\beta}(x)}\right)^{\!\!(0)}-\left(\frac{\delta R^{\gamma\delta}(y)  }{\delta g_{\alpha\beta}(x)}  \right)^{\!\!(0)}\right] 
  \underline{h}_{\gamma\delta}(y) \nonumber 
  \\
  &&\hspace{-1.2cm}
 +\left(-2\kappa_{D}^{-2}\right)\left(\gamma_{0}R^{(1)2}+\gamma_{2}R_{\mu\nu}^{(1)2}+\gamma_{4}R_{\mu\nu\rho\sigma}^{(1)2}\right) \, . 
\ee
\end{comment}

To be able to define the inverse of the kinetic operator 
we add to the Lagrangian the standard gauge-fixing term 
\[
{S}_{\rm gf}=-2\kappa_{D}^{-2}\int\! d^{D}x\left(-\frac{1}{\alpha}\right)\left(\partial^{a}h_{a\nu}-\beta\partial_{\nu}h_{a}^{a}\right)^{2} \, , 
\]
where we choose $\alpha=1$ and $\beta=1/2$, corresponding
to the usual harmonic gauge condition 
\begin{equation}
\partial^{a}h_{a\nu}=\frac{1}{2}\partial_{\nu}h^{a}_{a} \, .
\label{eq:HarmonicGauge}
\end{equation}
This is consistent with the physical conditions \eqref{onshellgrav} chosen for on-shell gravitons.
We can rewrite the kinetic operator in the momentum space as follows, 
\begin{equation}
\frac{1}{2}\underline{h}_{\alpha \beta}(-k)\left(\frac{\delta^{2}\left({S}_{g}+{S}_{\rm gf}\right)}{\delta g_{\alpha\beta}(- k)\delta g_{\gamma\delta}(k)}\right)^{\!\!(0)} 
\!\!\!
\underline{h}_{\gamma \delta}(k)
 \equiv \frac{1}{2}\underline{h}_{\alpha\beta}(-k)\mathcal{O}^{\alpha\beta,\gamma\delta}(k)\underline{h}_{\gamma\delta}(k) \, . 
\end{equation}
A standard procedure can be employed to compute the propagator $i \mathcal{O}^{-1}_{\alpha\beta,\gamma\delta}(k)$ as reviewed in \ref{appendixProp}. 

\subsection{The three and four-graviton vertices}
{\em Three graviton vertex ---}
The three graviton vertex is obtained by collecting all terms
in the action that are cubic in the graviton field $h_{\mu\nu}$.
Since we are interested in the on-shell four-point function, in particular
in the scattering amplitudes of four massless gravitons, we do not need
to go through a lengthy computation keeping track of all terms eventually
vanishing on-shell. 
Therefore, we can make full use of the simplification
brought about by the linearized vacuum equation of motion for the physical
field $h_{\mu\nu}$ in the harmonic gauge (\ref{eq:HarmonicGauge}), 
i.e. $\square h_{\mu\nu}=0$. 
Actually, our choice of polarizations
is such that we can assume the conditions \eqref{onshellgrav}
all along our computation, which greatly simplifies the algebra. 
In fact, these conditions imply that all the scalar operators are vanishing
on-shell at linear order in $h_{\mu\nu}$, including 
the scalar curvature $R^{\left(1\right)}$ and the root of metric determinant
$\sqrt{-g}^{\,\left(1\right)}$. One can further show that $R_{\mu\nu}^{\left(1\right)}=0$
due to the linearized EOM. The diagrams we need to compute are such
that for each three graviton vertex two gravitons out of three are on-shell and one 
is off-shell because it must be contracted with the propagator of the internal line of the diagram. This means we can consider the physical and off-shell
gravitons as different fields, even from the combinatorial point of
view. Hence the procedure is first to expand the action to the first order
in the off-shell field $\underline{h}_{\mu\nu}$ and then 
to the second order in the physical field $h_{\mu\nu}$. 

It is convenient to rewrite the action (\ref{gravityStelle})
in terms of the only two combinations of the $\gamma_\ell$ parameters
that appear in the propagator, namely 
\be 
{S}_{g}=-2\kappa_{D}^{-2}\int\! d^{D}x\,\sqrt{-g} \Big(R+\left(\gamma_{0}-\gamma_{4}\right)R^{2}+\left(  \gamma_{2}+4\gamma_{4}\right)R_{\mu\nu}^{2} +
\gamma_{4}\left(R_{\mu\nu\rho\sigma}^{2}-4R_{\mu\nu}^{2}+R^2 \right) 
\Big) \, . 
\label{actionRedef}
\ee
We note the last term is the famous Gauss-Bonnet density, which is topological in four dimensions whereas for generic higher dimensions it gives rise to vertices only.\\
For the Einstein-Hilbert action ${S}_{\rm EH}=-2\kappa_{D}^{-2}\int\! d^{D}x\sqrt{-g}R$
we simply have for the three graviton vertex:
\be
\label{threevertexEH}
 i\left(\frac{\delta{S}_{\rm EH}}{\delta g_{\mu\nu}}\right)^{\!\! \left(2\right)}  \!\!\!\!
=  i \left(-2\kappa_{D}^{-2}\right)\Bigg[ \sqrt{-g}\left(\frac{1}{2}g^{\mu\nu}R-R^{\mu\nu}\right)\Bigg]^{\left(2\right)}
 \!\!\!\! =  i\left(-2\kappa_{D}^{-2}\right)\left(\frac{1}{2}\eta^{\mu\nu}R^{\left(2\right)}-R^{\mu\nu\left(2\right)}\right) 
\ee
Similar expressions arise from the scalar curvature square
action  
\be
{S}^{\prime}_{0}= -2 \kappa_{D}^{-2} \left(\gamma_{0}-\gamma_{4}\right)\int\! d^{D}x\,\sqrt{-g}R^{2},
\ee 
the Ricci square action
\be
{S}^{\prime}_{2}= -2\kappa_{D}^{-2} \, \left(\gamma_{2}+4\gamma_{4}\right)\int\! d^{D}x\,\sqrt{-g}R_{\mu\varrho}^{2} \, , 
\ee
and the Gauss-Bonnet action 
\be 
{S}^{\prime}_{4}= -2\kappa_{D}^{-2}\, \gamma_{4}\int\! d^{D}x\,\sqrt{-g}\left(R_{\mu\nu\rho\sigma}^{2}-4R_{\mu\nu}^{2}+R^2\right) \, .
\ee
The outcome is:
\begin{eqnarray}
&& \hspace{-1cm}
 i\left(\frac{\delta {S}^{\prime}_{0}}{\delta g_{\mu\nu}}\right)^{\!\!\left(2\right)}\!\!\!\!  =  i\left(-2\kappa_{D}^{-2}\right)\left(\gamma_{0}-\gamma_{4}\right)\Bigg[2\sqrt{-g}\left(g^{\mu\nu}\nabla^{2}-\nabla^{\mu}\nabla^{\nu}+\frac{1}{4}g^{\mu\nu}R-R^{\mu\nu}\right)R\Bigg]^{ \left(2\right)} \nonumber \\
 && \hspace{0.62cm}
  =  2i\left(-2\kappa_{D}^{-2}\right)\left(\gamma_{0}-\gamma_{4}\right)\left(\eta^{\mu\nu}\square-\partial^{\mu}\partial^{\nu}\right)R^{\left(2\right)} \, , \label{threevertexR} \\
&& \hspace{-1cm}
i\left(\frac{\delta {S}^{\prime}_{2}}{\delta g_{\mu\nu}}\right)^{\!\! \left(2\right)} \!\!\!\! =  i\left(-2\kappa_{D}^{-2}\right)\left(\gamma_{2}+4\gamma_{4}\right)\Bigg[\sqrt{-g}\left(\frac{1}{2}g^{\mu\nu}R^{\kappa\lambda}R_{\kappa\lambda}-2R^{\mu\lambda}R^{\nu}\phantom{}_{\lambda}\right)\Bigg]^{\left(2\right)} \nonumber \\
 &&  \hspace{0.62cm}
 +i   \left(-2\kappa_{D}^{-2}\right)\left(\gamma_{2}+4\gamma_{4}\right)\Big[\sqrt{-g}\left(-g^{\mu\kappa}g^{\nu\lambda}\nabla^{2}-g^{\mu\nu}\nabla^{\kappa}\nabla^{\lambda}+g^{\mu\kappa}\nabla^{\lambda}\nabla^{\nu}+g^{\nu\lambda}\nabla^{\kappa}\nabla^{\mu}\right)R_{\kappa\lambda}\Big]^{(2)} \nonumber \\
 &&  \hspace{0.62cm}
 =  i\left(-2\kappa_{D}^{-2}\right)\left(\gamma_{2}+4\gamma_{4}\right)\left(-\eta^{\mu\kappa}\eta^{\nu\lambda}\square-\eta^{\mu\nu}\partial^{\kappa}\partial^{\lambda}+\eta^{\mu\kappa}\partial^{\lambda}\partial^{\nu}+\eta^{\nu\lambda}\partial^{\kappa}\partial^{\mu}\right)R_{\kappa\lambda}^{\left(2\right)} , \label{threevertexRicci} 
\end{eqnarray}
 where $\square$ is the flat d'Alambertian operator. In $D=4$, as far as the computation of vertices is concerned, we can
ignore the presence of $S^{\prime}_4$ because of  the Gauss-Bonnet  theorem. The terms in \eqref{threevertexEH}, \eqref{threevertexR} and \eqref{threevertexRicci} contributing to the three graviton
vertex with two legs on-shell are expressed in terms of
the following second order expansions
\begin{eqnarray*}
&& R^{\left(2\right)}  =  -\partial_{b}h_{ac}\partial^{c}h^{ab}+\frac{3}{2}\partial_{c}h_{ab}\partial^{c}h^{ab}\, ,\\
&&R_{\mu\nu}^{\left(2\right)} =  \frac{1}{2}\partial_{\mu}h^{ab}\partial_{\nu}h_{ab} 
 +h^{ab}\left(\partial_{b}\partial_{a}h_{\mu\nu}+\partial_{\mu}\partial_{\nu}h_{ab}-\partial_{b}\partial_{\mu}h_{\nu a}-\partial_{b}\partial_{\nu}h_{\mu a}\right)
  +\partial^{b}h_{\mu}\phantom{}^{a}\left(\partial_{b}h_{\nu a}-\partial_{a}h_{\nu b}\right) \,  .
\end{eqnarray*}

{\em Four-graviton vertex ---}
For the on-shell four-graviton amplitudes we only need the four-graviton vertex with all the gravitons on-shell.
Therefore, by the same arguments used for the three vertex, we can argue that
\begin{eqnarray}
&& i\left({S}_{\rm EH}\right)^{\left(4\right)} =  -2\kappa_{D}^{-2}\int\! d^{D}x\left(\sqrt{-g}R\right)^{\left(4\right)} \nonumber \\
  && \hspace{1.56cm}
  =  -2\kappa_{D}^{-2}\int\! d^{D}x\left(R^{(4)}+\left(\sqrt{-g}\right)^{(2)}R^{(2)}\right)  ,\nonumber \\
&& i \left( {S}^{\prime}_{0}\right)^{\left(4\right)}  =  -2\kappa_{D}^{-2} \left(\gamma_{0}-\gamma_{4}\right)\int\! d^{D}x\left(R^{(2)}\right)^{2}  , \nonumber \\
&& i\left({S}^{\prime}_{2}\right)^{\left(4\right)}  =  -2\kappa_{D}^{-2} \left(\gamma_{2}+4\gamma_{4}\right)\int\! d^{D}x\left(R_{\mu\nu}^{(2)}\right)^{2}  .
\end{eqnarray}
Apart from $\left(\sqrt{-g}\right)^{(2)}=-\frac{1}{2}h_{\mu\nu}h^{\mu\nu}$
the only new quantity appearing in the four vertex is $R^{(4)}$, whose expression is given in \ref{appendixExp}. We remind the derivation of all vertices from ${S}_{\rm EH}$, ${S}^{\prime}_{0}$ and ${S}^{\prime}_{2}$ is valid in any dimension $D$.

\begin{comment}
\begin{eqnarray*}
\left(\sqrt{-g}Rf_{scalar}\left(\nabla^{2}\right)R\right)^{\left(4\right)} & = & R^{\left(2\right)}f_{scalar}\left(\square\right)R^{\left(2\right)}\\
\left(\sqrt{-g}R^{\mu\nu}f_{scalar}\left(\nabla^{2}\right)R_{\mu\nu}\right)^{\left(4\right)} & = & R^{\mu\nu\left(2\right)}f_{scalar}\left(\square\right)R_{\mu\nu}^{\left(2\right)}
\end{eqnarray*}
\end{comment}

\section{The amplitudes in Stelle gravity}
In this section we explicitly evaluate the four-graviton scattering amplitudes for quadratic Stelle gravity using the method of Feynman diagrams. We distinguish the two cases of four and higher dimensions. In the first case we have already introduced all the ingredients which will be used in the computation. For the second one we will add new vertices and extend the form of polarization tensors. This second case will be the first extension of our results first obtained in quadratic four-dimensional Stelle theory.

\subsection{4D Stelle gravity}
\label{4DStelle}
\begin{figure}
\begin{center}
% h h \to h h
\begin{picture}(150,150)(0,0)
% lines of the diagram
\Line(20,10)(60,60)
\Line(130,10)(90,60)
\Line(20,140)(60,90)
\Line(130,140)(90,90)
% blob
\GCirc(75,75){22}{0.5}
% arrows
\LongArrow(20,30)(40,55)
\LongArrow(130,30)(110,55)
\LongArrow(60,110)(40,135)
\LongArrow(90,110)(110,135)
% arrows' descriptions
\Text(20,45)[l]{$p_1$}
\Text(132,45)[r]{$p_2$}
\Text(50,130)[l]{$p_3$}
\Text(102,130)[r]{$p_4$}
% polarizations' descriptions
\Text(30,10)[l]{$\varepsilon_{1\mu\nu}$}
\Text(125,10)[r]{$\varepsilon_{2\mu\nu}$}
\Text(10,125)[l]{$\varepsilon^*_{3\mu\nu}$}
\Text(145,125)[r]{$\varepsilon^*_{4\mu\nu}$}
\end{picture}
\caption{Kinematics of the process $hh \to hh$.}
\label{kinematics}
\end{center}
\end{figure}
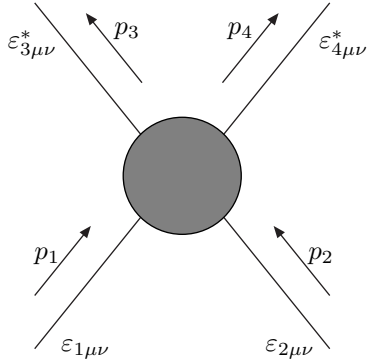

In Born approximation there are four diagrams to be considered, which include the contact one and the ones with virtual propagation in the $s$, $t$ and $u$ channels. They are depicted in Fig.\ref{fourgravscat}, while the kinematics of the process is showed in Fig.\ref{kinematics}.
\begin{figure}[h]
\begin{center}
% s channel
\begin{picture}(360,120)(0,0)
% lines of the diagram
\Line(10,10)(60,40)
\Line(110,10)(60,40)
\Line(60,40)(60,80)
\Line(10,110)(60,80)
\Line(110,110)(60,80)
% vertices
\GCirc(60,40){1}{0}
\GCirc(60,80){1}{0}
% description
\Text(42,0)[l]{$s$ channel}
%
%
% t channel
% lines of the diagram
\Line(130,10)(160,60)
\Line(230,10)(200,60)
\Line(160,60)(200,60)
\Line(130,110)(160,60)
\Line(230,110)(200,60)
% vertices
\GCirc(160,60){1}{0}
\GCirc(200,60){1}{0}
% description
\Text(162,0)[l]{$t$ channel}

% u channel
% lines of the diagram
\Line(250,10)(270,60)
\Line(350,10)(330,60)
\Line(270,60)(330,60)
\Line(250,110)(330,60)
% interrupted graviton's leg 
\Line(350,110)(306,83)
\Line(294,75)(270,60)
% vertices
\GCirc(270,60){1}{0}
\GCirc(330,60){1}{0}
% description
\Text(282,0)[l]{$u$ channel}
\end{picture}

\begin{picture}(120,120)(0,0)
% lines of the diagram
\Line(10,10)(60,60)
\Line(110,10)(60,60)
\Line(10,110)(60,60)
\Line(110,110)(60,60)
% vertex
\GCirc(60,60){1}{0}
% description
\Text(42,0)[l]{contact}
\end{picture}

\caption{ Tree-level Feynman diagrams giving rise to the process of four-graviton scattering.}
\label{fourgravscat}
\end{center}
\end{figure}
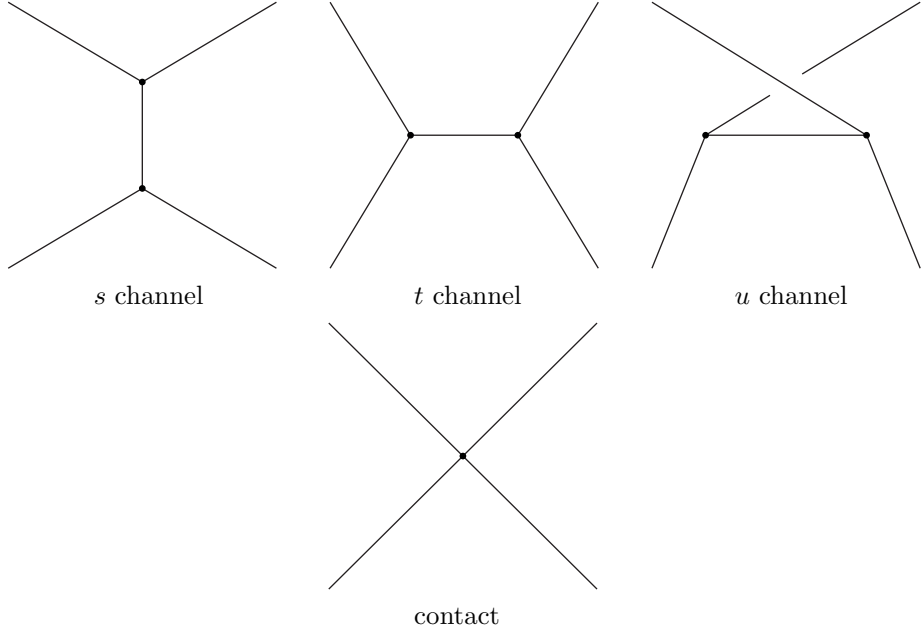

The exchange diagrams can be computed
by contracting two three graviton vertices with the propagator. Out of the
five possible tensor structures in the propagator (see \ref{appendixProp}) only three ($X_1$, $X_2$ and $X_3$) are seen
to have a non-vanishing contribution as a consequence of the fact that we are considering on-shell gauge invariant amplitudes. In order to be able to carry
out the algebra effectively, it's very convenient to express the Mandelstam
variables in terms of the energy $E$ and the scattering angle $\theta$ in the center-of-mass
reference frame, 
\be
s=4E^2 \, , \quad t=-2E^{2}\left(1-\cos\theta\right) \quad {\rm and} \quad 
u=-2E^{2}\left(1+\cos\theta\right) \, . 
\ee
With this choice of variables and assuming all the momenta lying in the plane defined by $\phi = 0$,  the incoming ($p_1$ and $p_2$) and outgoing ($p_3$ and $p_4$) gravitons are identified by the momenta \eqref{prototypemomentum} and polarizations \eqref{poltensor} obtained from \eqref{prototypepolarization}, where $\bar\theta$ is chosen to be $0$, $\pi$, $\theta$ and $\pi+\theta$ respectively. For outgoing gravitons the complex conjugates of polarization tensors should be used.

 The contributions from different diagrams to the amplitude for gravitons with all helicities $+2$ are
\be
&& \hspace{-1cm}
\mathcal{A}_{s}\left(++,++\right)  = -i \left(-2 \kappa_4^{-2} \right) \frac{9}{8} E^2 \sin ^2\theta 
 -i \left(-2 \kappa_4^{-2} \right) \frac{9}{4}  E^4 \Big[8(\gamma_0-\gamma_4)+(\gamma_2+4\gamma_4) \left(\cos 2 \theta +3\right)\Big] \, , \\
&& \hspace{-1cm}
\mathcal{A}_{t}\left(++,++\right)  =   -i\frac{\left(-2\kappa_{4}^{-2}\right)\cos^{4}\frac{\theta}{2}}{256(\cos\theta-1)}\Big[E^{2}(392\cos\theta+80\cos2\theta-8\cos3\theta-720) \\
 && \hspace{1.5cm}
  +(\gamma_0-\gamma_4)E^{4}\left(4\cos\theta+16\cos2\theta-6\cos3\theta-4\cos4\theta+2\cos5\theta-12\right) \nonumber \\
 && \hspace{1.45cm}
  +(\gamma_2+4\gamma_4)E^{4}\left(2146\cos\theta-216\cos2\theta-99\cos3\theta+6\cos4\theta+\cos5\theta-1838\right)\Big] \, ,  \nonumber \\
%%%%% 
&& \hspace{-1.0cm}
\mathcal{A}_{u}\left(++,++\right)  =  -i\frac{\left(-2\kappa_{4}^{-2}\right)\sin^{4}\frac{\theta}{2}}{256(\cos\theta+1)}\Big[E^{2}(392\cos\theta-80\cos2\theta-8\cos3\theta+720)\\
 && \hspace{1.5cm}  +(\gamma_0-\gamma_4)E^{4}\left(4\cos\theta-16\cos2\theta-6\cos3\theta+4\cos4\theta+2\cos5\theta+12\right)  \nonumber\\
 && \hspace{1.5cm}
 +(\gamma_2+4\gamma_4)E^{4}\left(2146\cos\theta+216\cos2\theta-99\cos3\theta-6\cos4\theta+\cos5\theta+1838\right)\Big]\, ,  \nonumber \\
 && \hspace{-1cm}
\mathcal{A}_{\rm contact}\left(++,++\right)  =  i\frac{\left(-2\kappa_{4}^{-2}\right)}{1024}\Big[E^{2}(160\cos2\theta-8\cos4\theta+872) \\
 && \hspace{2.2cm}+(\gamma_0-\gamma_4)E^{4}\left(-34\cos2\theta+4\cos4\theta+2\cos6\theta+18460\right)  \nonumber\\
 && \hspace{2.2cm}+(\gamma_2+4\gamma_4)E^{4}\left(1711\cos2\theta-46\cos4\theta+\cos6\theta+9598\right)\Big] \, . 
  \nonumber
\ee
These results themselves contain a nontrivial piece of information.
In fact, in such a covariant gauge as the harmonic one we have chosen, at any energy scale 
the propagator behaves like
\begin{equation}
\frac{1}{k^{2}\left(1-\gamma^{\prime}k^{2}\right)}=\frac{1}{k^{2}}-\frac{1}{k^{2}-\gamma^{\prime-1}}\label{eq:PropScale-1}
\end{equation}
and because of the small $k$ expansion $\left[k^{2}\left(1-\gamma^{\prime}k^{2}\right)\right]^{-1}=k^{-2}+\gamma^{\prime}-\gamma^{\prime2}k^{2}+\ldots$
we would expect quantities dependent on arbitrary powers of $E^{2}$
, whereas we find that only energies up to the fourth power show up and in particular no energy powers appear in the denominators. This is the sign of nontrivial cancellations happening between the three graviton vertices and propagators for each diagram.  The ghost poles in \eqref{eq:PropScale-1} disappear at any energy scale and moreover we are safe from IR divergences. However the contribution from each diagram is not separately gauge invariant and only the sum giving the full amplitude matters.
\begin{comment}
Is this fact related to the fact that Stelle theory is renormalizable
in 4D? From this point of view, it would be interesting to have the
full set of polarizations to check whether higher powers of the energy
appear in higher dimensions.
\end{comment}

Actually, it's easy to check that the full scattering amplitude reads
\begin{eqnarray*}
&& \mathcal{A}\left(++,++\right)  =  \mathcal{A}_{s}\left(++,++\right)+\mathcal{A}_{t}\left(++,++\right)+\mathcal{A}_{u}\left(++,++\right)+\mathcal{A}_{\rm contact}\left(++,++\right)\\
 &&\hspace{2.0cm} =  -2i\left(-\frac{2}{\kappa_{4}^2} \right)E^{2}\frac{1}{\sin^2\theta} \, ,
\end{eqnarray*}
which is the same result as the one for the tree-level amplitude
of Einstein theory. We can also find a similar result for the other
three independent helicity amplitudes
\be
&& \mathcal{A}\left(+-,+-\right)  =-\frac18i\left(-\frac{2}{\kappa_{4}^2} \right)E^{2}\frac{(1+\cos\theta)^4}{\sin^2\theta} \, ,\\
&&\mathcal{A}\left(++,+-\right)=\mathcal{A}\left(++,--\right)=0\, .
\ee
As a consequence of another nontrivial cancellations these results coincide with the very well known one \cite{EHcomputations} for tree-level graviton scattering
in Einstein theory in spite of the fact that the dimensional arguments
that completely constrain the form of helicity amplitudes in the latter
case \cite{Grisaru:1975bx} can no longer be naively applied to quadratic gravity in $D=4$,
because of the presence of new dimensionless couplings in the theory. We will find a natural explanation for these surprising results  in section \ref{fieldredefinition}.

\subsection{$D>4$ Stelle gravity}
\label{Dbigger4}
For $D>4$ the situation becomes more complicated. First of
all, the Gauss-Bonnet term is no longer topological and so it gives
a non-vanishing contribution to the three and four-graviton vertices.
Apart from the terms we have already computed, for ${S}_{4}=\left(-2\kappa_{D}^{-2}\right)\gamma_{4}\int\! d^{D}x\sqrt{-g}R_{\mu\nu\rho\sigma}^{2}$
we find
\begin{eqnarray}
&& \hspace{0cm}
i\left(\frac{\delta{S}_{4}}{\delta g_{\mu\nu}}\right)^{\left(2\right)} \!\!\!\! =  %\left(-2\kappa_{D}^{-2}\right)\gamma_{4}\left(2\sqrt{-g}\left(\frac{1}{4}g^{\mu\nu}R^{\kappa\lambda\rho\sigma}R_{\kappa\lambda\rho\sigma}-g^{\nu\tau}R^{\mu\lambda\rho\sigma}R_{\tau\lambda\rho\sigma}+ \delta R^{\mu}\phantom{}_{\nu\kappa\lambda}R_{\mu}\phantom{}^{\nu\kappa\lambda}\right)\right)^{(2)} \nonumber \\
% && =  
 \left(-2\kappa_{D}^{-2}\right)\gamma_{4}\Bigg[2\sqrt{-g}\left(\frac{1}{4}g^{\mu\nu}R^{\kappa\lambda\rho\sigma}R_{\kappa\lambda\rho\sigma}-g^{\nu\tau}R^{\mu\lambda\rho\sigma}R_{\tau\lambda\rho\sigma}+\nabla_{\lambda}\nabla_{\kappa}R{}^{(\mu|\lambda|\nu)\kappa}\right)\Bigg]^{(2)} \nonumber \\
% && =  \left(-2\kappa_{D}^{-2}\right)\gamma_{4}\left(\frac{1}{2}\eta^{\mu\nu}\left(R_{\kappa\lambda\rho\sigma}^{(1)}\right)^{2}-2\eta^{\nu\tau}\eta^{\mu\kappa}R_{\kappa\lambda\rho\sigma}^{(1)}R_{\tau\lambda\rho\sigma}^{(1)}+\left(\nabla_{\lambda}\nabla_{\kappa}R{}^{\mu\lambda\nu\kappa}\right)^{(2)}\right) \, ,
% \nonumber \\
  &&\hspace{1.7cm} =  \left(-2\kappa_{D}^{-2}\right)\gamma_{4}\Big[\frac{1}{2}\eta^{\mu\nu}\left(R_{\kappa\lambda\rho\sigma}^{(1)}\right)^{2}-2\eta^{\nu\tau}\eta^{\mu\kappa}R_{\kappa\lambda\rho\sigma}^{(1)}R_{\tau\lambda\rho\sigma}^{(1)}-\frac12\partial^\mu\partial^\nu R^{(2)}+\square R^{\mu\nu (2)} \Big]\, ,
 \nonumber \\
%\end{eqnarray}
%
%\begin{eqnarray}
&& \hspace{0.4cm}
i\left({S}_{4}\right)^{\left(4\right)}  =  \left(-2\kappa_{D}^{-2}\right)\gamma_{4}\int\! d^{D}x\Big[\left(R_{\mu\nu\rho\sigma}^{(2)}\right)^{2}+2R_{\mu\nu\rho\sigma}^{(1)}R_{\mu\nu\rho\sigma}^{(3)} +8\,g^{\tau\lambda\, (1)}R_{\tau\nu\rho\sigma}^{(1)}R_{\lambda\nu\rho\sigma}^{(2)}\nonumber \\
 && \hspace{1.7cm} 
 +12\,g^{\tau\lambda\,(1)}g^{\kappa\nu\,(1)}R_{\tau\kappa\rho\sigma}^{(1)}R_{\lambda\nu\rho\sigma}^{(1)}+4\,g^{\tau\lambda\, (2)}R_{\tau\nu\rho\sigma}^{(1)}R_{\lambda\nu\rho\sigma}^{(1)}
  +\sqrt{-g}^{\,(2)}\left(R_{\kappa\lambda\rho\sigma}^{(1)}\right)^{2}\Big] \, .
\end{eqnarray}
The three and four-graviton vertices are therefore determined in terms of the on-shell non-vanishing quantities, whose explicit expressions are given in \ref{appendixExp}. As we will see from the results below, it is crucial to include the term $\sqrt{-g}R_{\mu\nu\rho\sigma}^{2}$ in the action in higher dimensions. This term is naturally motivated by effective field theory considerations and addition of it is in another direction, how we can extend four-dimensional theory.

An additional complication is given by the fact that in $D>4$ the
graviton has $\frac{D\left(D-3\right)}{2}$ polarizations. This is
a trivial consequence of the fact that the little group for a massless
particle in $D$ dimensions is $SO(D-2)$ and that the graviton is
by definition identified with the traceless symmetric rank 2 tensor representation.
Because a massless vector particle with momentum $p_\mu$ ($p^{2}=0$) has
$(D-2)$ independent polarizations $\epsilon_{\mu}^{D}(p,\lambda)$
we can choose the Lorentz gauge condition $p^{\mu}\epsilon_{\mu}=0$ and identify $\epsilon_{\mu}$ up to a gauge transformation 
$\epsilon_{\mu}\rightarrow\epsilon_{\mu}+\gamma\, p_{\mu}$ (for any $\gamma$) to single out
an irreducible representation of $SO(D-2)$ with $\lambda$ identifying
the elements of the vector basis. This basis can be conveniently chosen
as the one such that
\[
H^{i}\epsilon_{\mu}^{D}(p,\lambda)=\lambda^{i}\epsilon_{\mu}^{D}(p,\lambda) ,
\]
with $H^{i}$ the elements of the Cartan subalgebra , $i=1,\ldots,r$.
For $SO(D-2)$ we can break up the $(D-2)$-dimensional space
into $\left[{D}/{2}\right]-1$ different two-dimensional subspaces. 
The rotation operator $W^{k}$ ($k=1,\ldots, \left[{D}/{2}\right]-1$) acts on the subspace $(2k,  2k-1)$.
Therefore polarization vectors are identified by the $D-2$ weights
of the fundamental representation of $SO(D-2)$. If we assume the momentum $p_\mu$
to be in the $(D-1)$-th spatial direction and $W^{\left[D/2 \right]-1}$
to be the generator of rotations in the $(D-3$, $D-2)$ plane, then
we should consider in particular the polarizations corresponding to the $\left[D/2 \right]-1$ dimensional weights
$\left(0,\ldots,0,\pm1\right)$, i.e. the $D$-vector $\left(0,\ldots,0,1,\pm i,0\right)$. In the following computation we will concentrate on this structure of $D$-dimensional polarization vectors. Now we want to repeat the discussion in the section \ref{helicityamp} for a general dimension $D$.

For a generic momentum whose spatial components are only in $(D-3,\, D-2,\, D-1)$-subspace
\be p_{\mu}=\left(p_{0},0,\ldots,0,p_{0}\sin\bar\theta\cos\phi,p_{0}\sin\bar\theta\sin\phi,p_{0}\cos\bar\theta\right), 
\ee
we have 
\[
\epsilon_{\mu}^{D}(p,\pm)=\left(0,\ldots,0,\cos\bar\theta\cos\phi\mp i\sin\phi,\cos\bar\theta\sin\phi\pm i\cos\phi,-\sin\bar\theta\right)\,.
\]
These polarizations can be used to construct two traceless symmetric
tensors satisfying the gauge condition $p^{\,\mu}\epsilon_{\mu\nu}=0$ and equivalent up to a gauge transformation
$\epsilon_{\mu\nu}\rightarrow\epsilon_{\mu\nu}+a_{\mu}p_{\nu}+a_{\nu}p_{\mu}$, where
$a\cdot p=0$,
\[
\epsilon_{\mu\nu}^{D}(p,\pm)=\epsilon_{\mu}^{D}(p,\pm)\epsilon_{\nu}^{D}(p,\pm) \,.
\]
Gravitons in $D$ dimensions, which we consider here, are with two possible polarizations $+$ or $-$, similarly to the case in four dimensions.
%
%
\begin{comment}
pol A is the $\epsilon_{\mu\nu}^{+}=\epsilon_{\mu}^{+}\epsilon_{\nu}^{+}$
and $\epsilon_{\mu}^{+}$ is the $4D$ $\epsilon_{\mu}^{+}$ polarization
with extra 0 in the extra spatial dimensions.

pol B is the $\epsilon_{\mu\nu}=\epsilon_{\mu}\epsilon_{\nu}$ and
$\epsilon_{\mu}^{+}$ is a vector with a $1$ and an $i$ in the two
extra spatial dimension and $0$ everywhere else.
\end{comment}
%

In this framework we have carried out the computation for the amplitude
$\mathcal{A}^{D}\left(++,++\right)$ in $D=5$ and $6$ dimensions. These are the results
\be
&& \hspace{-1.3cm}
\mathcal{A}^{D=5}\left(++,++\right)=-i\frac{2}{\kappa_{5}^{2}} \left\{ \frac{16E^{6}\gamma_{4}^{2}\left[1+8E^{2}\left(3(\gamma_{0}-\gamma_{4})+(\gamma_{2}+4\gamma_{4})\right)\right]}{\left(1-4E^{2}(\gamma_{2}+4\gamma_{4})\right)\left[3+4E^{2}\left(16(\gamma_{0}-\gamma_{4})+5(\gamma_{2}+4\gamma_{4})\right)\right]}
-2    %\left(-2\kappa_{D}^{-2}\right)
E^{2}\frac1{\sin^{2}\theta} \label{5Dreszero}
\right\} ,  \\
&& \hspace{-1.3cm}
\mathcal{A}^{D=6}\left(++,++\right)=
-i\frac{2}{\kappa_{6}^{2}}
\left\{ 
\frac{8E^{6}\gamma_{4}^{2}\left[1+8E^{2}\left(3(\gamma_{0}-\gamma_{4})+(\gamma_{2}+4\gamma_{4})\right)\right]}{\left(1-4E^{2}(\gamma_{2}+4\gamma_{4})\right)\left[1+2E^{2}\left(10(\gamma_{0}-\gamma_{4})+3(\gamma_{2}+4\gamma_{4})\right)\right]}
-2 
%\left(-2\kappa_{D}^{-2}\right)
E^{2}\frac1{\sin^{2}\theta}
\right\} .
\label{6Dreszero}
\ee
Quite beautifully, 
each amplitude is the sum of two terms, one comes from the 
usual Einstein gravity, whereas the other one is
overall proportional to $\gamma_{4}^{2}$. This implies that in the
absence of the Riemann square action ${S}_{4}$ the (quite boring) result
found in four dimensions holds true in higher dimensions too, making the
scalar curvature square term ${S}_{0}$ and the Ricci square
term ${S}_{2}$ undetectable in the tree-level graviton scattering amplitudes. This is why the presence of the $S_4$ term in the action is crucial in higher dimensions. The mechanism of cancellation in the former case is the same as in four dimensions.

Another interesting feature is that, apart from the standard Einstein term, the remaining dependence of the amplitudes on the parameters of the quadratic terms
is only through the combinations $\gamma_{0}-\gamma_{4}$ and $\gamma_{2}+4\gamma_{4}$. 
To better understand we first define a new quadratic action for gravity with parameters $\gamma'_0$, $\gamma'_2$ and $\gamma_4$ according to 

\be 
{S}_{g}=-2\kappa_{D}^{-2}\int\! d^{D}x\,\sqrt{-g} \Big(R+\gamma'_{0}R^{2}+  \gamma'_{2}R_{\mu\nu}^{2} +
%\underbrace{ 
\gamma_{4}{\rm GB}  
\Big) \, ,
\label{actionredefined}
\ee
which is completely equivalent to \eqref{gravityStelle}, and where GB denotes the Gauss-Bonnet Lagrangian. We find precisely that $\gamma'_0$ and $\gamma'_2$ are equal to combinations appearing in \eqref{5Dreszero} and \eqref{6Dreszero}. Therefore those results can be rewritten in a more compact form
\be
&& \hspace{-1.3cm}
\mathcal{A}^{D=5}\left(++,++\right)=-i\frac{2}{\kappa_{5}^{2}} \left\{ \frac{16E^{6}\gamma_{4}^{2}\left[1+8E^{2}(3\gamma'_{0}+\gamma'_{2})\right]}{\left(1-4E^{2}\gamma'_{2}\right)\left[3+4E^{2}(16\gamma'_{0}+5\gamma'_{2})\right]}
-2    %\left(-2\kappa_{D}^{-2}\right)
E^{2}\frac1{\sin^{2}\theta} \label{5Dres}
\right\} ,  \\
&& \hspace{-1.3cm}
\mathcal{A}^{D=6}\left(++,++\right)=
-i\frac{2}{\kappa_{6}^{2}}
\left\{ 
\frac{8E^{6}\gamma_{4}^{2}\left[1+8E^{2}(3\gamma'_{0}+\gamma'_{2})\right]}{\left(1-4E^{2}\gamma'_{2}\right)\left[1+2E^{2}(10\gamma'_{0}+3\gamma'_{2})\right]}
-2 
%\left(-2\kappa_{D}^{-2}\right)
E^{2}\frac1{\sin^{2}\theta}
\right\} .
\label{6Dres}
\ee

The dependence on the parameters $\gamma_i$ in the two above formulas can be easily explained diagrammatically. We know that the propagator derived from \eqref{actionredefined} doesn't depend on $\gamma_4$ coefficient in any $D$, only cubic and quartic vertices derived from the Gauss-Bonnet term possess such dependence (precisely expressions for them are linear in the $\gamma_4$ parameter). The propagator depends only on $\kappa_D^{-2}$, $\gamma'_0$ and $\gamma'_2$. Hence our conclusion is that besides contribution to amplitudes \eqref{5Dres} and \eqref{6Dres} from Einstein gravity, we have additional contributions from exchange diagrams, where both three graviton vertices are derived from the Gauss-Bonnet term and on the internal line we have the full propagator of the theory. The dependence on $\gamma'_0$ and $\gamma'_2$ is only through the propagator, but not through vertices. The dependence on $\kappa_D^{-2}$ is determined by dimensional reasons. The fact, that there is no any linear dependence on $\gamma_4$ in the final results forces us to believe that here we are witnessing another cancellation between contact diagram with vertex from Gauss-Bonnet term and exchange diagrams with two different vertices (one from GB, the second one from standard terms $R$, $R^2$ or $R^2_{\mu\nu}$). This interpretation of the results is quite natural, because we have modified the theory (with already many cancellations) only by addition of new vertices coming from the Gauss-Bonnet term, but the propagator has remained the same. We have also computed the amplitudes for other choices of polarizations
and checked that their general properties are similar.

Moreover, as it is obvious from \eqref{5Dres} and \eqref{6Dres} the new terms in the amplitudes are associated with the appearance of arbitrary powers of ${E}^{2}$
in the expansion of the denominators in the infrared regime, while in the ultraviolet regime the highest power is of course $E^4$, because this is a quadratic gravity. This result already extends the previous findings \cite{Hochberg:1986hj, Deser:1986xr}, which were restricted only to the order $E^4$ in energy expansion around $E=0$. When we treat the theory \eqref{actionredefined} as fundamental, then we do not need to focus on the low energy limit of the amplitudes and we have the exact energy dependence in the tree-level amplitudes. By comparing the cross sections computed in \cite{Hochberg:1986hj}, we could in principle read out the values of the parameters $\gamma'_0$, $\gamma'_2$, and $\gamma_4$. Then it would be natural to associate the parameter $\gamma_4$ with the strength of Gauss-Bonnet interactions, 
while two other  $\gamma'_0$ and $\gamma'_2$ would be related to the masses of the ghost and the curvaton, which appear in the spectrum of quadratic gravity. However, this theory is not unitary (due to the presence of the ghost) and we cannot conclude about new physically meaningful contributions to the graviton scattering compared to the amplitudes computed in Einstein gravity. In the next section we would like to address the same issue in a more realistic theory.

\section{Four-graviton scattering amplitudes in nonlocal gravity}  
\label{sorpresa}
In this section we explicitly calculate the four-graviton scattering amplitudes 
for the weakly nonlocal theory  
(\ref{gravity})
with zero potential $({\bf V} =0)$, namely 
\be
&& 
\mathcal{L}_{\rm g} = -  2 \kappa_{D}^{-2} \, \sqrt{-g} 
\left[ R 
+
R \, 
 \gamma_0(\square)%_{\Lambda})
 R 
 +  R_{\mu\nu}
\gamma_2(\square)%_{\Lambda})
 R^{\mu\nu}
 + R_{\mu\nu\rho\sigma}  \, 
\gamma_4(\square)
R^{\mu\nu\rho \sigma}
\right]  .
\label{gravityV0}
\ee
This is another direction of extension of our original four-dimensional result in Stelle gravity from \ref{4DStelle}. Here we assume the theory to be valid in any dimension $D$, but the explicit formulas will be given for the case $D=4$ (in higher dimensions only the numerical coefficients change, but the final results are the same). Additionally results from this section can be used in particular for any local higher derivative theory, whose action is quadratic in gravitational curvature and contains a finite number of derivatives. We remark that for below results to hold true the assumptions about super-renormalizability (or only renormalizability), unitarity and nonlocality of the theory are \emph{not} by any means necessary. Since we compute the tree-level scattering these issues are irrelevant. We decide to speak about the theory \eqref{gravity}, because this is a candidate theory to have a good behavior at quantum level too and could be viewed as a fully consistent realization of quantum gravity in quantum field theory framework.

The stunning result is once again the same of Einstein gravity for the case $\gamma_4(\square) = 0$ in any dimension. 
Since on-shell the expansion of two the simplest curvature invariants is ${\bf R}\sim O(h^2)$ and ${\bf Ric}\sim O(h^2)$, the form factors are spectators in the expansion in number of gravitons and many of the results of the previous section still apply to the general nonlocal theory \eqref{gravityV0}. In more technical terms, when we compute the variations to find propagator and vertices, we don't need to vary the form factors and original covariant boxes $\square$ being the argument thereof. Putting all the $n$-point functions on flat spacetime we can substitute these arguments with flat d'Alembertian operators. Finally after going to momentum space we can easily replace them with negative invariant squares of the momenta -- Mandelstam variables -- $s$, $t$ and $u$ for each channel respectively \cite{Eran}. Only for the contact diagrams we have to be careful and we closely investigate on which graviton legs the operator inside the form factor acts to properly associate dependence on $s$, $t$ and $u$ in form factors in momentum space.

 We here report the result in $D=4$ and with form factors appearing only inside curvature scalar and Ricci tensor squared terms (no terms with Riemann tensors sandwiching the form factor $\gamma_4(\square)$) for the amplitude $\mathcal{A}(++,++)$ and we 
 explicitly show the cancellations leading to the important result announced above. 
 The amplitude $\mathcal{A}(++,++)$ gets contributions from the contact diagram and ones with graviton exchanges
 in the $s$, $t$ and $u$ channels (compare Fig.\ref{fourgravscat}.), namely 
\be
&& \hspace{-1cm}
\mathcal{A}_s(++,++) =  -2\kappa_4^{-2} \, \left(-\frac{9}{8}\frac{t(s+t)}{s}
{+\frac{9}{32} \gamma_2(s)\left(s^{2}+(s+2t)^{2}\right)}
+{\frac{9}{8} s^{2} \gamma_0(s)} \right) ,
\label{sC} \\
%%%
&&\hspace{-1cm}
 \mathcal{A}_t(++,++)=  -2\kappa_4^{-2} \, \left(-\frac{1}{8}\frac{\left(s^{3}-5s^{2}t-st^{2}+t^{3}\right)(s+t)^{2}}{s^{3}t} \right.\nonumber \\ 
&& \hspace{1.4cm}
\left.+\frac{1}{16}\gamma_2(t)\frac{\left(2s^{4}-10s^{3}t-s^{2}t^{2}+4st^{3}+t^{4}\right)(s+t)^{2}}{s^{4}}
  +\frac{1}{8}  \gamma_0(t)\frac{t^{2}(s+t)^{4}}{s^{4}}\right)  ,
  \label{tC}
\\
&& \hspace{-1cm}
\mathcal{A}_u(++,++)=  -2 \kappa_4^{-2} \, \left(-\frac{1}{8}\frac{\left(s^{3}-5s^{2}u-su^{2}+u^{3}\right)(s+u)^{2}}{s^{3}u} \right.\nonumber \\
 && \hspace{1.5cm}
 \left. +\frac{1}{16} \gamma_2(u)\frac{\left(2s^{4}-10s^{3}u-s^{2}u^{2}+4su^{3}+u^{4}\right)(s+u)^{2}}{s^{4}}
  +\frac{1}{8} \gamma_0(u)\frac{u^{2}(s+u)^{4}}{s^{4}} \right)  ,
  \label{uC} \\
%\ee
%%%
%\be
&&  \hspace{-1cm}
\mathcal{A}_{\rm contact}(++,++)=  -2 \kappa_4^{-2} \, \left(-\frac{1}{4}\frac{s^{4}
+s^{3}t - 2st^{3}-t^{4}}{s^{3}} \right. %\nonumber \\
 %&&
 { -\frac{9}{32}\gamma_2(s)\left(s^{2}+(s+2t)^{2}\right)}
 {-\frac{9}{8}s^{2}\gamma_0(s)}\nonumber \\
 && \hspace{2.2cm}
 { -\frac{1}{16}\gamma_2(t)\frac{\left(2s^{4}-10s^{3}t-s^{2}t^{2}+4st^{3}+t^{4}\right)(s+t)^{2}}{s^{4}}
 }
 { -\frac{1}{8}\gamma_0(t)\frac{t^{2}(s+t)^{4}}{s^{4}} }\nonumber  \\
 && \hspace{2.2cm}
 \left.{
 -\frac{1}{16}\gamma_2(u)\frac{\left(2s^{4}-10s^{3}u-s^{2}u^{2}+4su^{3}+u^{4}\right)(s+u)^{2}}{s^{4}}
 }
 {-\frac{1}{8}\gamma_0(u)\frac{u^{2}(s+u)^{4}}{s^{4}} } \right)  .
 \label{cC}
\ee
The full amplitude is given by the sum of above contributions (\ref{sC})$+$(\ref{tC})$+$(\ref{uC})$+$(\ref{cC}), and the result is:
\be
\hspace{-0.4cm}
\boxed{ \mathcal{A}(++,++) = \mathcal{A}_s(++,++) + \mathcal{A}_t(++,++) + \mathcal{A}_u(++,++) + \mathcal{A}_{\rm contact}(++,++) = \mathcal{A}(++,++)_{\rm EH} }\, ,
 \ee
 where $\mathcal{A}(++,++)_{\rm EH}$ is the amplitude for the Einstein-Hilbert theory. Notice that all terms but the first of equation \eqref{sC} cancel with the last two terms of the first line of \eqref{cC}. Analogously terms from \eqref{tC} cancel with the second line of \eqref{cC} and those from \eqref{uC} cancel with the last line of \eqref{cC}.
For the other helicity amplitudes (\ref{helicity}) we get analogous simplifications.

Once the operator quadratic in Riemann tensor (or Weyl tensor, or the generalized Gauss-Bonnet term) is 
turned on the scattering amplitudes change radically.  
Not only amplitudes will explicitly depend on the form factor $\gamma_4(\square)$, but also 
on the other form factors in the theory, namely $\gamma_0(\square)$ and $\gamma_2(\square)$, as it is evident from the five- and six-dimensional results reported in \eqref{5Dres}, \eqref{6Dres} for the case of constant form factors. There $\gamma_4$ is an overall factor,
and if it does not vanish, the results will depend also on $\gamma_0$ and $\gamma_2$. However the mechanism for acquiring such dependences is exactly the same as described in section \ref{Dbigger4}.  The case of computation with non-vanishing form factor $\gamma_4(\square)$ is qualitatively different from presented here. In this situation to obtain vertices we need to vary the form factor and covariant boxes in it. The reason for this is simple to explain -- the expansion of Riemann tensor on-shell starts at the first order, ${\bf Riem}\sim O(h)$, which is in opposition to the case for Ricci tensor and curvature scalar. 

When the nonlocal generalized Gauss-Bonnet operator 
\be
{\rm GB}_{\gamma_4(\square)}={\bf Riem} \, \gamma_4(\square) {\bf Riem}-4{\bf Ric} \, \gamma_4(\square) {\bf Ric}+{\bf R} \, \gamma_4(\square) {\bf R}
\label{GGB}
\ee
is switched on the amplitudes 
will depend explicitly on $\gamma_4(\square)$ (through vertices) and on $\gamma'_0(\square)=\gamma_0(\square)-\gamma_4(\square)$ and $\gamma'_2(\square)=\gamma_2(\square)+4\gamma_4(\square)$ (through propagators).
This operator is non-trivial in any dimension $D\geq4$, and gives rise only to new vertices, but not to the full propagator of the theory. As it is now obvious from the discussion in section \ref{Dbigger4} the vertices derived from this term contain derivatives of the form factor with respect to its argument (up to the second order). Moreover the argument of the form factor, which is the covariant box operator (acting on a tensor field up to rank four) must be varied too (up to the second order), which complicates the situation quite a lot. Therefore the computation of such vertices is quite involved and we don't attempt to present the exact results here. We only remark that the results for constant form factors in section \ref{Dbigger4} for $D>4$ are consistent with these qualitatively described here. The presence of a non-constant form factor $\gamma_4(\square)$ in a nonlocal theory \eqref{gravityV0} is crucial even in $D=4$ and there in an ideal graviton scattering experiment we could fit the form factors $\gamma'_0(\square)$, $\gamma'_2(\square)$ and $\gamma_4(\square)$ by measuring the cross section.

\section{General $n-$graviton scattering amplitudes in local and nonlocal theories}
\label{fieldredefinition}
Having discussed the scattering amplitudes in four-dimensional Stelle theory and in simple extensions of it 
in the previous sections (higher dimensional setup, inclusion of terms quadratic in Riemann tensor, and nonlocal form factors) now we wish to explore other possible directions about how we could extend our results. The main motivation is to seek for a well-defined setup in which amplitudes will differ significantly from those obtained in Einstein theory and therefore would permit for unambiguous verification of the theory by comparing predictions for cross sections with hypothetical experiments on gravitational wave scattering. As we will see in this section it is still difficult to depart from very ubiquitous results of Einstein gravity. As a first step we will try to describe the situation when  
$n$-point correlation functions are considered. In this area the theorem (first proven by Anselmi in \cite{Anselmi:2002ge, Anselmi:2006yh}) will reveal to be very enlightening for the tree-level situation. We will again find typically only standard results from Einstein theory and a justification for this will be given. A special role will be assigned to the Riemann tensor and to the form factor $\gamma_4(\square)$. At the end we will comment on other possible extensions by the inclusion of other operators: local (and higher in curvatures) and in a sense more nonlocal, and by going beyond tree-level. In this way we will touch on any reasonable extension of the original theory \eqref{gravityStelle}. Exploration of all these directions in an exhaustive way is beyond the scope of this paper and we will leave it for future publications.

Let us consider the case of higher than four-graviton scattering amplitudes at tree level. For this goal we present here the following theorem. 

\noindent Theorem. {\em All the $n$-point functions in any gravitational theory $($in particular super-renormalizable or finite$)$ with an action } 
\be
&& 
\mathcal{L}_{\rm gr} = -  2 \kappa_{D}^{-2} \, \sqrt{-g} 
\left[ {\bf R} 
+
{\bf R} \, 
 \gamma_0(\square)%_{\Lambda})
 {\bf R} 
 + {\bf Ric} \, 
\gamma_2(\square)%_{\Lambda})
 {\bf Ric} 
+ {\bf V({\bf R}, {\bf Ric}}, {\bf Riem}, \nabla) \,
\right]  , 
\label{gravitySFnoRiem}
\ee
{\em give the same on-shell tree-level amplitudes as the Einstein-Hilbert theory, 
$ \mathcal{L}_{\rm EH} = -  2 \kappa_{D}^{-2} \, \sqrt{-g} \, {\bf R }$, provided that the potential ${\bf V}$ 
is at least quadratic in ${\bf Ric}$ and/or {\bf R}. In particular for any theory in which we can recast the potential in the 
following form 
\be
{\bf V} = {\bf Ric} \cdot {\bf \tilde{V}} \cdot {\bf Ric} \equiv R_{\mu\nu} [{\bf \tilde V({\bf R}, {\bf Ric}, {\bf Riem}, \nabla)  }]^{\mu\nu\rho\sigma} R_{\rho \sigma},
\label{VRic}
\ee
the theorem is still valid} (${\bf \tilde V  }$ {\em is in full generality a differential operator with contravariant indices $\mu,\nu,\rho,\sigma$ acting on the Ricci tensor to the right, containing at least one power of gravitational curvature.$)$}

Proof. The proof is based on the field redefinition theorem proved by Anselmi \cite{Anselmi:2006yh}
at perturbative level and to all orders in the Taylor expansion of the redefinition of the metric field.

First we assume that we have given two general weakly nonlocal action functionals $S'(g)$  and $S(g')$, respectively defined in terms of the metric fields 
$g$ and $g'$, such that
%we first modify $S'$ into 
\be
S'(g) = 
%S(g') 
 S(g) + E_i(g) F_{i j}(g)  E_j (g) \, ,
 \label{AnselmiC}
\ee
where $F$ can contain derivative operators and ${E_i = \delta S/\delta g_i}$ is the EOM of the theory with action $S(g)$\footnote{Here we use a compact deWitt notation and with the indices $i$, $j$ on fields we encode all Lorentz, group indices, and the spacetime dependence of the fields. Additionally, we assume that the field space is flat and we do not need to raise indices in sums there.}. The statement of the theorem is that there exists a field redefinition 
\be
g_i'  = g_i + \Delta_{i j} E_j  \quad \Delta_{i j} = \Delta_{j\hspace{0.03cm} i}, 
\ee
such that, perturbatively in $F$, but to all orders in powers of $F$, we have the equivalence
\be
S'(g) = 
S(g')  \,.
\label{FR}
%= S(g) + E_i F^{i j}  E_j (g),
\ee
Above $\Delta_{ij}$ is a possibly nonlocal operator acting linearly on the EOM $E_j$, with indices $i$ and $j$ in the field space, and it is defined perturbatively %in terms of
in powers of the operator $F_{ij}(g)$, namely $\Delta_{ij}=F_{ij}(g)+\ldots$
% starting at, or infinitesimally
Let us consider the 
%It is easy to prove the theorem at the 
first order in the Taylor expansion for the functional $S(g')$, which reads
\be
S(g') = S(g + \Delta g) \approx S(g) + \hspace{-0.05cm}  %\overbrace{ 
\frac{\delta S}{\delta g_i} %}^{ {\rm EOM} }
\hspace{-0.05cm}   \delta g_i = 
 S(g) + E_i   \, \delta g_i \, .
\ee
If we can find a weakly nonlocal expression for $\delta g_i$ such that 
\be
S'(g) = %S(g') \approx 
S(g) + E_i   \, \delta g_i 
\ee
(note that the argument of the functionals $S'$ and $S$ is now the same), 
then there exists a field redefinition $g\rightarrow g'$ satisfying \eqref{FR}. Hence the two actions $S'(g)$ and $S(g')$ are tree-level equivalent.
\vspace{-0.65cm}
\begin{flushright}
$\square$    
\end{flushright}
\vspace{-0.2cm}

As it is obvious from above, in the proof of our theorem it was crucial to use classical EOM $E_i$. In the theory \eqref{gravitySFnoRiem} this implies ${\bf Ric}=0$ in vacuum regions without the presence of any matter source. These are the conditions we have imposed on the linearized level when we defined asymptotic states of on-shell gravitons as perturbative states of our theory.

Now we can apply the above field redefinition theorem to our class of theories \eqref{gravity}, where we do not  include terms with Riemann tensor $\bf Riem$. Since we are interested in $S(g') \equiv S_{\rm EH}(g')$ and $S'(g) \equiv S_{\rm gr}(g)$, the relation \eqref{AnselmiC} reads
\be
%S(g') \equiv 
%S_{\rm EH}(g') \equiv
 S(g') = S_{\rm EH} (g) + R_{\mu \nu}(g)  F^{\mu \nu, \rho\sigma}(g)  R_{\rho\sigma}(g)
%\stackrel{=}{gg}
= S'(g) 
%\equiv S_{\rm gr}(g)
\,.
\ee
Here we also used that in the spectrum of Einstein-Hilbert theory we only have the massless spin 2 graviton (contrary to the case of the theory described by the action \eqref{gravitySFnoRiem} with polynomial form factors) and the on-shell scattering of such particles we relate in the two theories. It would be clearly 
nonsensical to apply the theorem for scattering of other particles (appearing e.g. in theory \eqref{gravitySFnoRiem}) and attempt to relate it to the scattering amplitudes in the Einstein two-derivative theory.

If the potential ${\bf V}$ is at least quadratic in ${\bf R}$ and/or ${\bf Ric}$, namely takes the form \eqref{VRic},
%\be
%{\bf V} = {\bf Ric} \cdot {\bf \tilde{V}} \cdot {\bf Ric} \equiv R_{\mu\nu} 
%{ [V({\bf R}, {\bf Ric}, {\bf Riem}, \nabla)]  }^{\mu\nu\rho\sigma} R_{\rho %\sigma},
%\ee
where ${\bf \tilde{V}}$ is a rank four tensor made of any tensor including Riemann tensor, its contractions and derivative operators,
then the tensor $F_{ij}(g)$ used in the field redefinition \eqref{AnselmiC} exists, it is weakly nonlocal, and equals to 
\be
%\hspace{-0.5cm}
F^{\mu\nu, \rho\sigma} =  g^{\mu\nu} g^{\rho\sigma} \gamma_0(\square) 
+ g^{\mu\rho} g^{\nu\sigma} \gamma_2(\square) +
%+ (x_1 R^{\mu\nu}  R^{\rho\sigma} + x_2 \, g^{\mu\nu}  g^{\rho\sigma} + 
%x_3 \, R^{\mu\nu}  g^{\rho\sigma} +{\rm perm.}) 
{\bf \tilde V({\bf R}, 
{\bf Ric}, {\bf Riem}, \nabla)  } ^{\mu\nu\rho\sigma}  \, .
\ee
In the additional part of our theorem we will discuss about the possibility of recasting the potential $\bf V$ in the form \eqref{VRic}. Of course due to the known ambiguity related to an order of writing covariant derivatives (non-commuting on a general manifold) in tensorial expressions, this last remark is not very precise. More precisely we require that the contribution to the EOM from the potential $\bf V$ should vanish, when the ansatz ${\bf Ric}=0$ is used. (This is the conclusion about vacuum spacetime in Einstein gravity, which on the linearized level coincides with the on-shell conditions for the massless gravitons.) We do not say that in order to use the theorem the initial potential $\bf V$ must be in the form \eqref{VRic}, it only must be possible to cast it in such form.

Our results about Stelle theory in four dimensions reported in section \ref{4DStelle} can be understood in the following way. First, we may employ Gauss-Bonnet theorem to reduce the action to the form with only the Ricci scalar and the Ricci tensor square terms. Secondly, we can use the field redefinition theorem to prove that this theory is %perturbatively 
equivalent to the Einstein-Hilbert theory regarding the on-shell tree-level graviton amplitudes. Another convenient choice is to start with a theory written in a Weyl basis with the quadratic part consisting of the Ricci scalar square ${\bf R}^2$ and the Weyl tensor square ${\bf C}^2$. We can take the limit in which only the coefficient in front of ${\bf C}^2$ survives so that the theory is now conformally invariant at classical level for $D=4$.
%Now in this theory we can perform the reduction again to Einstein theory exploiting some facts about conformal gravity given in $D=4$ by the Weyl Lagrangian ${\bf C}^2$, which is conformally invariant at classical level. 
The operator ${\bf C}^2$ should contribute with the fourth power in the energy to the four-graviton scattering amplitudes, but conformal invariance requires the scattering amplitudes to be numbers independent on the scale. Therefore, the amplitudes must be zero because the graviton field is dimensionless and there is no other scale in Weyl gravity. 
On the other hand, in $\mathcal{N}=4$ super-Yang-Mills theory we can have non-zero amplitudes because the gauge bosons have energy dimension one. 
However, we explicitly proved that also the operator ${\bf R}^2$ does not give any contribution to the amplitudes, so that we can conclude the scattering amplitudes for any purely quadratic gravity in $D=4$ are vanishing.

As a special case, our explicit computation confirms that the four-graviton scattering amplitudes in
four dimensional Weyl {\em conformal gravity} \cite{man} is identically zero, namely 
\be
\mathcal{L} = - \alpha_g \sqrt{-g} \, C_{\mu\nu\rho \sigma} C^{\mu\nu\rho \sigma} \quad \Longrightarrow  \quad 
\mathcal{A}(\mbox{4-graviton}) \equiv 0 \, .
\ee
% is {\em identically zero}.
%This is why it is tree-level CFT, and in such theories scattering amplitudes must vanish for dimensionless fluctuations in $D=4$ and without the possibility of using any dimensionful parameter (there is no such in the theory). 
This derives from the fact that in any CFT $S$ matrix must be trivial.

In $D>4$, operators quadratic in the Riemann tensor (with or without form factors) cannot be recast in the form of 
operators quadratic in the Ricci and scalar curvatures without introducing extra vertices 
because the generalized Gauss-Bonnet operator (\ref{GGB}) is no more topological. Therefore, they could contribute to the scattering amplitudes. % Moreover, in $D\geqslant 6$ the {\bf Riem} will be present also in the vertices 
%(Ex. ${\bf Ric}^2 \cdot {\bf Riem}$). 
This completely explains the results found in section \ref{Dbigger4} and also explain why 
in the case of form factors sandwiched among Riemann tensors the results will deviate from the ones computed in Einstein theory. In the four dimensional case the generalized Gauss-Bonnet term (\ref{GGB}) does have an impact on the EOM and the Ricci-flatness ansatz is not valid in the vacuum of the theory. Therefore, also in this case we cannot apply the theorem and the amplitudes differ from those of the Einstein theory in agreement with the results reported in section \ref{sorpresa}. In particular the amplitudes in the Einstein-Hilbert action supplemented by the two-loop correction term (Goroff-Sagnotti term \cite{Goroff:1985th}), understood as effective field theory, will not coincide with those from pure Einstein theory.

We comment also on applications of the above theorem to the case of dimension $D$.
For the case of a finite theory the result can be extended to any order in the loop expansion if we neglect the finite contributions to the quantum action. In even dimension we can easily achieve finiteness by using of killers constructed at least out of two Ricci tensors (compare formula \eqref{Minimal} for $D=4$). Note that in odd dimension, for $\gamma>(D-1)/2$ in the description after formula \eqref{Tomboulis}, the theory \eqref{gravityV0} is finite in DIMREG without the need to add any killer operator. We also note that in $D=3$ the Riemann tensor is not independent from the Ricci tensor and the scalar, %{\bf Ric} and {\bf R}
therefore three-dimensional Einstein-Hilbert gravity without matter is finite at quantum level (but without  perturbative degrees of freedom). 

For the case of a 1-loop super-renormalizable theory in $D=4$ the theorem can be applied at any order in the loop expansion including quantum loop divergences of amplitudes and the quantum logarithmic corrections coming together with the one-loop running of gravitational couplings.
However, we again expect deviations from Einstein Hilbert amplitudes 
due to other quantum finite contributions. It is crucial here that the theory is one-loop super-renormalizable, because we only have divergences at the controllable one-loop level. In a general renormalizable theory we would have divergences of the type $R^2$ and $R_{\mu\nu}^2$ at any loop order in $D=4$, 
and the structure of the RG equations for the running coupling constants or
the finite terms in the effective action will be much more complicated. In the case of theories renormalizable and (super-renormalizable) in higher dimension $D\geq6$ we cannot apply our theorem any more, because then operators of the type ${\bf Riem}^3$ in the potential ${\bf V}$ are needed for having a renormalizable theory.

To make the discussion of the redefinition theorem more %clear and 
transparent we present here an example of its use.

{\em Example.} For the finite four-dimensional theory \eqref{Minimal}, the following choice of $F$ makes \eqref{Minimal}  tree-level equivalent to the Einstein-Hilbert theory (all the graviton scattering amplitudes are the same):
\be
F^{\mu\nu, \rho\sigma} = g^{\mu\rho} g^{\nu\sigma} \frac{e^{H(-\square_{\Lambda})}-1}{\square} 
- \frac{1}{2} g^{\mu\nu} g^{\rho\sigma} \frac{e^{H(-\square_{\Lambda})}-1}{\square} 
+s_1 g^{\mu\nu} g^{\rho\sigma} R \square (R \cdot)
+s_2 R^{\mu\nu} \square (R^{\rho\sigma} \cdot)  \, .
\ee
 (When the action of the field redefinition (contraction $F^{\mu\nu, \rho\sigma}R_{\rho\sigma}$) is evaluated, we substitute the center dot in the last two terms above by the Ricci tensor on the right $R_{\rho\sigma}$.)

{\em Remark.} We here showed that $S_{\rm gr}$ and $S_{\rm EH}$ are tree-level %pertubatively 
equivalent 
and all the on-shell scattering amplitudes can be equivalently calculated using one or the other theory.
However, off-shell amplitudes do not match, because in proving the theorem we made crucial use 
of the equations of motion ${\bf Ric}=0$, that uniquely characterize the perturbative graviton field in vacuum in both 
the theories. Similarly in the domain of classical field theory other non-perturbative solutions exist in $S_{\rm gr}$, which are not shared by $S_{\rm EH}$. Therefore the two theories 
are equivalent only in the framework of perturbation theory, and full matching of the amplitudes happens only at tree-level.

As reviewed in section \ref{FINITA}, we can introduce other operators, local or nonlocal, to make the theory \eqref{gravity} finite at quantum level. In this section making use of the 
field redefinition theorem we proved 
that the killer operators in the action \eqref{Minimal} do not give tree-level contributions to the $n-$graviton on-shell scattering amplitudes. However, \eqref{Minimal} is not unique and we are free to introduce other killers that can affect the $n$-point functions. 
One example of such killers is the following quartic operator in the Riemann tensor, 
\be
s_4 \, {\bf Riem}^2 \square^{\gamma -2} {\bf Riem}^2, 
\label{RiemK}
\ee
which gives contribution to the four and $n-$graviton scattering amplitudes ($n\geq4$),
as shown in the proof above, %indeed from this operator 
because we can extract from it at least four-graviton fields around flat spacetime  
(we remind that we get ${\bf Riem} \sim O(h)$, while ${\bf Ric} \sim O(h^2)$ when expanding the metric around Minkowski spacetime with on-shell metric fluctuations).

{\em Beyond tree-level amplitudes}. 
Finally, we expand about the operators we expect
 %on the possible expectations of the form of he operators 
beyond the simplest tree-level computation. In this short subsection the main emphasis is placed on the finite terms and their contributions to the scattering amplitudes at loop levels. Of course it is known that such terms are not universal and the results for the amplitudes are unambiguous only if some renormalization conditions at some energy scale are fixed (below it is assumed that is is already done). 

At quantum level Einstein gravity is non-renormalizable and we expect contributions to $n-$graviton amplitudes from many other operators unlike the case of 1-loop super-renormalizable or finite theories. 
For the latter we expect to have an upper limit on the number of derivatives in the UV for the operators in quantum effective action (precisely $2 \gamma + 4$ in $D=4$),
while for Einstein gravity we formally have up to infinite number of them.

At one-loop a super-renormalizable theory in $D=4$ gets extra nonlocal 
contributions that in the UV look like 
\be
%&& 
{\bf R} \log \left( - \frac{\square}{\mu^2} \right) {\bf R} \, , \quad %\nonumber \\
%\,\,\,\,\, {\rm and }  \,\,\,\,\, && 
{\bf Ric} \log \left( - \frac{\square}{\mu^2} \right)  {\bf Ric} \, ,
\label{RRicQ}
\ee
and by virtue of the results in this paper they do not contribute to any graviton scattering amplitudes. 
Conversely the finite contribution, which arises in higher dimensions,
\be
{\bf Riem} \log \left( - \frac{\square}{\mu^2} \right) {\bf Riem} \, ,
\ee
will give an explicit or implicit (after recasting in terms of the operators (\ref{RRicQ}) plus extra vertices) contribution.
Moreover, we have an upper limit for the UV energy scaling of the scattering amplitudes evaluated with the quantum action. 
For the super-renormalizable theory without Riemann tensors, or a finite theory with killers not involving them,  the upper limit is $ E^4$ in $D=4$, while in extra dimensions it is $ E^{D}$. %For a finite theory with Killers and/or terminators.

% jacobian
With the field redefinition at hand we may attempt to compute the contribution even beyond tree-level in theories for which we can still apply our theorem. In this situation the two theories differ only by the Jacobian of the transformation that we need to include when we compute the partition function with the path integral method. 
As we proved above the action functionals are equivalent, i.e. $S(g')=S'(g)$. In the path integral we also have to be careful about the change of the measure -- Jacobian of the transformation. The Jacobian of the field redefinition transformation $g \to g'$ is given by
\be
{\cal J} = {\rm Det}\frac{\delta g}{\delta g'}  = {\rm Det}\frac{\delta g_{\mu\nu}(x)}{\delta g'_{\rho\sigma}(x')} = {\rm Tr}\, \log  \frac{\delta g_{\mu\nu}(x)}{\delta g'_{\rho\sigma}(x')}%= {\rm Tr}\, {\rm Ln} \frac{\delta g_{\mu\nu}(x)}{\delta g'_{\mu\nu}(x)}
\,.
\ee
Its divergent part vanishes in DIMREG scheme for the case of any 
analytic field redefinition.
However, the finite non-analytic contributions to the quantum action, if any, take part in the Jacobian that  
we can compute perturbatively. For this purpose we introduce two ghost-like auxiliary fields (similar to Faddeev-Popov ghosts), which have fermionic statistics \cite{Marcus:1984ei}. Next, we can consider Feynman loop diagrams with these fields and we compute the contribution of the Jacobian to the scattering amplitudes. 
At the zero-loop order (tree-level) the Jacobian is one and we should do not worry about the inclusion of diagrams with the new ghosts in the amplitudes. The contribution of the Jacobian starts at the one-loop order.

In $D=5$ we may expect the following finite terms in the effective action,
\be
  q_0 \, {\bf R} \sqrt{   - \square  }  \, {\bf R} \,,  \quad %\nonumber \\ % \,\,\,\,
 q_2 \, {\bf Ric} \sqrt{   -\square  } \, {\bf Ric} \,,  \quad %\nonumber \\ %
 q_4 \,  {\bf Riem} \sqrt{   -\square  }\, {\bf Riem} \, .
 \label{sqrt}
 \ee
The non-analytic functions (like the above square root) of the covariant box operator appear due to dimensional reasons in any odd dimension. Only the last operator with two Riemann tensors will contribute to the amplitudes, and the coefficient $q_4$ will keep track 
of all the other form factors present in the classical or quantum action as explicitly evaluated in section \ref{sorpresa}.

We do not know at the moment the other finite contributions to the quantum action, but we know, as already mentioned, what the upper limit for the energy scaling of the finite contributions to the quantum action is: $E^4$ in $D=4$. In the high energy regime ($E\gg \kappa_D^{-1} \sim M_P \sim \Lambda$) we expect these
finite contributions to approach the form of the following operators:
\be
&& {\mathcal{R}} \frac{1}{\square} {\mathcal{R}} \frac{1}{\square} {\mathcal{R}}^2 \, , \nonumber \\%\, , \,\,\,\, 
&&  {\mathcal{R}} \frac{1}{\square} {\mathcal{R}} \frac{1}{\square}{\mathcal{R}} \frac{1}{\square} {\mathcal{R}}^2 \, , \nonumber \\%\, , \,\,\,\,
&&
{\mathcal{R}} \frac{1}{\square} {\mathcal{R}} \frac{1}{\square} {\mathcal{R}}\frac{1}{\square}  \dots  \frac{1}{\square} {\mathcal{R}}^2 \, , \nonumber \\
&& \quad  \dots \, , 
\label{Rbox-1}
\ee
which probably contribute to the scattering amplitudes. 

Anyhow, only an explicit calculation of the
one-loop scattering amplitudes will tell us the form of such corrections.

\section{Conclusions}
In this paper we have performed a tree-level computation of on-shell four-graviton scattering amplitudes in the context of higher derivative Stelle theories and nonlocal gravitational theories quadratic in the curvature with non-locality  specified by form factors. The theories in the first class are known to be generically non-unitary due to the appearance of non-physical poles in the spectrum \cite{Stelle}. Conversely, the second class of theories,  under some specific choice of the form factors, have been proven to be good candidates for a ghost-free and super-renormalizable or finite theory of quantum gravity \cite{modesto,modestoLeslaw, Briscese, Krasnikov, Tombo}. In both cases we have checked that, in the absence of terms quadratic in the Riemann tensor, the amplitudes coincide with the ones found in Einstein theory. 
Furthermore, the four-graviton scattering amplitudes in Weyl conformal gravity are identically zero. 

We provided an explanation of this result on the basis of a field redefinition map of quadratic gravity into the usual Einstein-Hilbert action \cite{Anselmi:2006yh}. This map is defined perturbatively at all orders in the parameters appearing in quadratic gravity. This ensures the two theories, sharing the same unperturbed action, are completely equivalent from the point of view of the tree-level amplitudes. 
More specifically this equivalence holds only as long as the two theories are supposed to have the same free spectrum. We cannot use the map to address the computation of observable quantities involving poles not present in Einstein theory. 

The idea that field redefinitions do not affect the physical $S$ matrix was actually discussed in the context of quantum field theory, specifically renormalization theory, a long time ago %\cite{Marcus:1984ei} 
and it actually lies at the core of the fundamental results about the quantum divergences of Einstein gravity 
\cite{'tHooft:1974bx, Goroff:1985th}. An application of the field redefinition very similar to the one in this paper
can be found in \cite{Deser:1986xr}. In \cite{Deser:1986xr} the authors argue that, just thanks to field redefinition, the order $\alpha^\prime k^4$ four-graviton amplitudes calculated in string theory can be derived from an effective action, where only the Gauss-Bonnet density shows up, consistently with  the previous predictions in \cite{Zwiebach:1985uq}. The field redefinition was also used to prove that in any higher derivative effective gravitational action the effective graviton propagator is always without ghosts. Similar results are derived at large in \cite{Anselmi:2002ge}. Our calculation is actually very near in the spirit to the one in \cite{Hochberg:1986hj}, where the effective quadratic action,  
reproducing the string theory amplitudes at the order $\alpha^\prime k^4$, is determined by only considering the $O(k^4)$ amplitudes. However, in the approximation considered, the presence of the additional poles in the propagator contributes only  
linearly in $\gamma_0$ and $\gamma_2$. 

In this paper we have considered the full non-linear dependence including the full propagator of quadratic gravity and  checked the redefinition theorem to all orders in $\gamma_0$ and $\gamma_2$.  When the Riemann square term contributes to the interaction vertices (in $D>4$) we have found additional terms in the amplitudes, depending on $\gamma_0$ and $\gamma_2$, which were neglected in previous computations. This is justified because the previous computations were to order $O(k^4)$, while the first non-vanishing vertices' contribution are possible to be derived from terms at least cubic in curvature (so at least of order $O(k^6)$).
Furthermore, the on-shell four-graviton amplitudes for a large class of weakly nonlocal gravity theories have been computed drawing on the fact  that the form factors present in the action can be treated without much effort
 when on-shell gravitons are considered \cite{Piva, Eran}. We emphasize that our results differ from the standard ones obtained in Einstein gravity for the case of Gauss-Bonnet term in higher dimensions and also in the case of generalized Gauss-Bonnet terms (sandwiching a function of the d'Alembertian $\Box$ operator (\ref{GGB})) in $D\geq4$. In principle, this allows us to determine the form factors by comparing the results for amplitudes with hypothetical experimental data on graviton scattering. In particular we are able to determine the coefficients in the most general quadratic Stelle theory in $D=6$ dimensions. Among other extensions of our results we mentioned amplitudes with any number of external gravitons ($n\geq4$) and our expectations about results beyond the tree-level.

Finally, although the original motivation of this study was to evaluate scattering amplitudes in a 
particular class of weakly nonlocal theories of gravity, the outcome of the paper is a general 
feature of any higher derivative local or nonlocal gravitational theory: Einstein quantum gravity \cite{Anselmi:2002ge}, conformal gravity, effective string theory, and local 
or nonlocal higher derivative super-renormalizable theories 
 \cite{shapiro3}. 
 We are now motivated to introduce more advanced techniques in order to calculate 
 $n-$point functions in local or nonlocal gravity involving the Riemann or Weyl tensors \cite{cachazo}.

\section*{Acknowledgements}
We are grateful to D. Anselmi and M. Piva for very useful discussions on the topic of field redefinitions in quantum field theory.

\appendix
\section{Details on the propagator}\label{appendixProp}
\noindent
In this appendix we  will closely follow the procedure employed in \cite{HigherDG}. 
Lorentz covariance and Bose symmetry allow us to rewrite the kinetic operator $\mathcal{O}_{\alpha\beta,\gamma\delta}(k)$ in \eqref{propagator}
making use of the following basis,  
\begin{eqnarray}
\label{PropOperators}
&& X_{1}  =  \eta_{\mu\nu}\eta_{\rho\sigma} \, ,  \quad 
X_{2}  =  \frac{1}{2}\left(\eta_{\mu\rho}\eta_{\nu\sigma}+\eta_{\mu\sigma}\eta_{\nu\rho}\right) \,  , \quad 
 X_{3}  =  \eta_{\mu\nu}\frac{k_{\rho}k_{\sigma}}{k^{2}}+\eta_{\rho\sigma}\frac{k_{\mu}k_{\nu}}{k^{2}} \, , \\
&& X_{4}  =  \frac{1}{4}\left(\eta_{\mu\rho}\frac{k_{\nu}k_{\sigma}}{k^{2}}+\eta_{\mu\sigma}\frac{k_{\nu}k_{\rho}}{k^{2}}+\eta_{\nu\rho}\frac{k_{\mu}k_{\sigma}}{k^{2}}+\eta_{\nu\sigma}\frac{k_{\mu}k_{\rho}}{k^{2}}\right)
\, , \quad 
X_{5}  =  \frac{k_{\mu}k_{\nu}k_{\rho}k_{\sigma}}{\left(k^{2}\right)^{2}} \, .
\end{eqnarray}
%Actually it turns out
We note that 
%\be
$\mathcal{O}_{\alpha\beta,\gamma\delta}(k)=\sum_{i=1}^{5}f_{i}X_{i}$
%\ee
does not depend on $\gamma_{0}$, $\gamma_{2}$, and $\gamma_{4}$, 
but only on the two linear combinations $\gamma_{0}-\gamma_{4}$ and
$\gamma_{2}+4\gamma_{4}$, because the integral
\be 
\int d^D x\,\left(R_{\mu\nu\rho\sigma}^{(1)2}-4R_{\mu\nu}^{(1)2}+R^{(1)2}\right)
\ee
is identically zero in any dimension $D$. From action \eqref{gravityV0} we read the coefficients $f_i$:
\be
&& \hspace{-0.8cm}
f_{1}  =  \left(-2\kappa_{D}^{-2}k^{2}\right)\left[\frac{1}{4}+2k^{2}\left(\left(\gamma_{0}-\gamma_{4}\right)+\frac{1}{4}\left(\gamma_{2}+4\gamma_{4}\right)\right)\right] \, , \quad 
 f_{2}  =  \left(-2\kappa_{D}^{-2}k^{2}\right)\left(-\frac{1}{2}\right)\left[1-k^{2}\left(\gamma_{2}+4\gamma_{4}\right)\right]  , \nonumber  \\
&& \hspace{-0.8cm}
f_{3}  =  \left(-2\kappa_{D}^{-2}k^{2}\right)\left(-2k^{2}\right)\left[\left(\gamma_{0}-\gamma_{4}\right)+\frac{1}{4}\left(\gamma_{2}+4\gamma_{4}\right)\right]  , \quad %\\&& \hspace{-0.8cm}
f_{4}  =  \left(-2\kappa_{D}^{-2}k^{2}\right)\left(-k^{2}\right)\left(\gamma_{2}+4\gamma_{4}\right)  ,  \nonumber \\
&& \hspace{-0.8cm}
f_{5}  =  \left(-2\kappa_{D}^{-2}k^{2}\right)k^{2}\left[2\left(\gamma_{0}-\gamma_{4}\right)+\left(\gamma_{2}+4\gamma_{4}\right)\right] \,.
\ee
In order to define the graviton propagator we have to invert the kinetic operator $ \mathcal{O}$. These calculations are most conveniently carried out in
terms of the Barnes-Rivers operators \cite{Rivers, VN} in the space of symmetric rank-two
tensors. The complete set of $D$-dimensional operators is given by
\begin{eqnarray*}
&& \hspace{-1.0cm}
P_{\mu\nu,\rho\sigma}^{1}  =  \frac{1}{2}\left(\theta_{\mu\rho}\omega_{\nu\sigma}+\theta_{\mu\sigma}\omega_{\nu\rho}+\theta_{\nu\rho}\omega_{\mu\sigma}+\theta_{\nu\sigma}\omega_{\mu\rho}\right) \, , \quad 
P_{\mu\nu,\rho\sigma}^{2}  =  \frac{1}{2}\left(\theta_{\mu\rho}\theta_{\nu\sigma}+\theta_{\mu\sigma}\theta_{\nu\rho}\right)-\frac{1}{D-1}\theta_{\mu\nu}\theta_{\rho\sigma} \, , \\
&&\hspace{-1.0cm}
P_{\mu\nu,\rho\sigma}^{0\theta}  =  \frac{1}{D-1}\theta_{\mu\nu}\theta_{\rho\sigma} \, , \quad 
P_{\mu\nu,\rho\sigma}^{0\omega}  =  \omega_{\mu\nu}\omega_{\rho\sigma} \, , \quad %\\&&\hspace{-1.0cm}
\bar{P}_{\mu\nu,\rho\sigma}^{0}  
=  P_{\mu\nu,\rho\sigma}^{0\theta\omega} + P_{\mu\nu,\rho\sigma}^{0\omega\theta} \, , \\
&& \hspace{-1.0cm}
P_{\mu\nu,\rho\sigma}^{0\theta\omega}  =  \frac{1}{\sqrt{D-1}}\theta_{\mu\nu}\omega_{\rho\sigma} \, , \quad 
P_{\mu\nu,\rho\sigma}^{0\omega\theta}  =  \frac{1}{\sqrt{D-1}}\omega_{\mu\nu}\theta_{\rho\sigma} \, , 
\end{eqnarray*}
where $\theta_{\mu\nu}$ and $\omega_{\rho\sigma}$ are the usual
transverse and longitudinal vector projection operators
\[
\theta_{\mu\nu}=\eta_{\mu\nu}-\frac{k_{\mu}k_{\nu}}{k^{2}},\quad
\omega_{\mu\nu}=\frac{k_{\mu}k_{\nu}}{k^{2}} \, , 
\]
that satisfy the relations: $\theta_{\mu\rho}\theta_{\sigma}^{\rho}=\theta_{\mu\nu}$, 
$\omega_{\mu\rho}\omega_{\nu}^{\rho}=\omega_{\mu\nu},$ and 
$\theta_{\mu\rho}\omega_{\nu}^{\rho}=0$.
%\be
%\theta_{\mu\rho}\theta_{\sigma}^{\rho}=\theta_{\mu\nu} \, , \quad \omega_{\mu\rho}\omega_{\nu}^{\rho}=\omega_{\mu\nu} \, , \quad
%\theta_{\mu\rho}\omega_{\nu}^{\rho}=0.
%\ee 
The operators $P^{1}$, $P^{2}$,$P^{0\theta}$ and $P^{0\omega}$
are idempotent, mutually orthogonal, and satisfy the following completeness relation, 
\[
\left[P^{1}+P^{2}+P^{0\theta}+P^{0\omega}\right]_{\mu\nu,\rho\sigma}=\frac{1}{2}\left(\eta_{\mu\rho}\eta_{\nu\sigma}+\eta_{\mu\sigma}\eta_{\nu\rho}\right)\equiv I_{\mu\nu,\rho\sigma} \, . 
\]
They project out the spin-1, spin-2, and two spin-0 parts of the field.
The two spin-0 transfer operators are such that
\be
&& \bar{P}^{0}P^{1}=P^{1}\bar{P}^{0}=\bar{P}^{0}P^{2}=P^{2}\bar{P}^{0}=0\, , \quad 
  \left(\bar{P}^{0}\right)^{2}=P^{0\theta}+P^{0\omega} \, , \quad \\
&&  P^{0\omega}\bar{P}^{0}=\bar{P}^{0}P^{0\theta}=P^{0\omega\theta}\, , \quad 
 P^{0\theta}\bar{P}^{0}=\bar{P}^{0}P^{0\omega}=P^{0\theta\omega}\, .
\ee
We decompose the operator $\cal O$ in the projectors basis 
\be
\mathcal{O}_{\alpha\beta,\gamma\delta}(k)=c_{1}P^{1}+c_{2}P^{2}+c_{0}^{\omega}P^{0\omega}+c_{0}^{\theta}P^{0\theta}+\bar{c}_{0}\bar{P}^{0},
\ee
where the coefficients 
$c_{1}, c_{2}, c_{0}^{\omega}, c_{0}^{\theta}, \bar{c}_{0}$ 
are obtained using the tensorial identities
\be
&& X_{1}  =  \left(D-1\right)P^{0\theta}+P^{0\omega}+\sqrt{D-1}\bar{P}^{0}\, , \quad 
X_{2}  =  P^{1}+P^{2}+P^{0\theta}+P^{0\omega} \nonumber \\
&& X_{3}  =  \sqrt{D-1}\bar{P}^{0}+2P^{0\omega}\, , \quad 
X_{4}  =  \frac{1}{2}P^{1}+P^{0\omega}\, , \quad 
X_{5}  =  P^{0\omega} \, . 
\ee
%%%
The coefficients $c_i$ are explicitly:
\be
&& \hspace{-1.3cm}
c_{1}  =  f_{2}+\frac{1}{2}f_{4} 
  =  \left(-2\kappa_{D}^{-2}k^{2}\right)\frac{1}{2}\, \quad 
c_{2}  =  f_{2} 
 = 2\kappa_{D}^{-2}k^{2} \left(1-k^{2}\left(\gamma_{2}+4\gamma_{4}\right)\right)\, , 
 \nonumber \\
&&  \hspace{-1.3cm}
c_{0}^{\theta}  =  \left(D-1\right)f_{1}+f_{2} %\, , \quad 
  =  \left(-2\kappa_{D}^{-2}k^{2}\right)\frac{1}{4}
  \left[ \left(D-3\right)+2k^{2}\left(4\left(D-1\right)\left(\gamma_{0}-\gamma_{4}\right)+D\left(\gamma_{2}+4\gamma_{4}\right)\right) \right] \,  ,\\
&&  \hspace{-1.3cm}
c_{0}^{\omega}  =  f_{1}+f_{2}+2f_{3}+f_{4}+f_{5}  =  \left(-2\kappa_{D}^{-2}k^{2}\right)\left(-\frac{1}{4}\right)
\, , \quad 
 \bar{c}_{0}  =  \sqrt{D-1}\left(f_{1}+f_{3}\right) 
  =  \left(-2\kappa_{D}^{-2}k^{2}\right)\frac{1}{4}\sqrt{D-1} \, . 
  \nonumber 
\ee
The problem of finding $\mathcal{O}_{\alpha\beta,\gamma\delta}^{-1}(k)=s_{1}P^{1}+s_{2}P^{2}+s_{0}^{\omega}P^{0\omega}+s_{0}^{\theta}P^{0\theta}+\bar{s}_{0}\bar{P}^{0}$
boils down to solving the linear system 
\[
\mathcal{O} \cdot  \mathcal{O}^{-1} = 
\begin{pmatrix}c_{1} & 0 & 0 & 0 & 0\\
0 & c_{2} & 0 & 0 & 0\\
0 & 0 & c_{0}^{\theta} & 0 & \bar{c}_{0}\\
0 & 0 & \bar{c}_{0} & 0 & c_{0}^{\omega}\\
0 & 0 & 0 & c_{0}^{\omega} & \bar{c}_{0}\\
0 & 0 & 0 & \bar{c}_{0} & c_{0}^{\theta}
\end{pmatrix}\begin{pmatrix}s_{1}\\
s_{2}\\
s_{0}^{\theta}\\
s_{0}^{\omega}\\
\bar{s}_{0}
\end{pmatrix}=\begin{pmatrix}1\\
1\\
1\\
0\\
1\\
0
\end{pmatrix}
 .
\]
Using the echelon matrix form \cite{HigherDG}
\[
\begin{pmatrix}c_{1} & 0 & 0 & 0 & 0 & 1\\
0 & c_{2} & 0 & 0 & 0 & 1\\
0 & 0 & c_{0}^{\theta} & 0 & \bar{c}_{0} & 1\\
0 & 0 & \bar{c}_{0} & 0 & c_{0}^{\omega} & 0\\
0 & 0 & 0 & c_{0}^{\omega} & \bar{c}_{0} & 1\\
0 & 0 & 0 & \bar{c}_{0} & c_{0}^{\theta} & 0
\end{pmatrix}\sim\begin{pmatrix}c_{1} & 0 & 0 & 0 & 0 & 1\\
0 & c_{2} & 0 & 0 & 0 & 1\\
0 & 0 & c_{0}^{\theta} & 0 & \bar{c}_{0} & 1\\
0 & 0 & 0 & c_{0}^{\omega} & \bar{c}_{0} & 1\\
0 & 0 & 0 & 0 & c_{0}^{\theta}c_{0}^{\omega}-\bar{c}_{0}^{2} & -\bar{c}_{0}\\
0 & 0 & 0 & 0 & 0 & 0
\end{pmatrix} \, ,
\]
%Therefore, t
the propagator is given by
\[
\mathcal{O}^{-1}=\frac{1}{c_{1}}P^{1}+\frac{1}{c_{2}}P^{2}+\frac{1}{c_{0}^{\theta}c_{0}^{\omega}-\bar{c}_{0}^{2}}\left(c_{0}^{\omega}P^{0\theta}+c_{0}^{\theta}P^{0\omega}-\bar{c}_{0}\bar{P}^{0}\right) \, .
\]

\section{Useful expansions}
\label{appendixExp}
\noindent
We list the expansions that are necessary to determine the propagator
and vertices used in the paper. The expansion of the metric around
flat spacetime is defined as 
\be
g_{\mu\nu}=\eta_{\mu\nu}+h_{\mu\nu} ,
\ee
%where $\eta_{\mu\nu}=\mathrm{diag}\left(-1,1,1,1\right)$. 
We will
always assume $h_{\mu\nu}$ as an on-shell field satisfying the conditions
$\square h_{\mu\nu}=0$, $\partial^\mu h_{\mu\nu}=0$, $h^\mu_\mu=0$. This is convenient
to reduce the expressions to a compact form which is the one we actually
need in most computations for on-shell gravitons amplitudes. When
we want to refer to unconstrained off-shell fields we will adopt the
notation $\underline{h}_{\mu\nu}$. 
\be
&& g^{\mu\nu(1)}=-h^{\mu\nu}\, , \nonumber \\
&& g^{\mu\nu(2)}=h^{\mu a}h_{a}{}^\nu \, , \nonumber \\
&& \left(\sqrt{-g}\right)^{(2)}  =  -\frac{1}{2}h_{ab}h^{ab} \, , \nonumber \\
&& R^{(1)} =  \partial_{a}\partial_{b}\underline{h}^{ab}-\square\underline{h}_{a}^{a} \, , \nonumber \\
&& R^{\left(2\right)}  =  -\partial_{b}h_{ac}\partial^{c}h^{ab}+\frac{3}{2}\partial_{c}h_{ab}\partial^{c}h^{ab}  \, , 
 \nonumber
 \ee
 \vspace{-.8cm}
 \be
 &&R^{\left(4\right)} = -12 h^{ab} h^{cd} \partial_{b}h_{de} \partial_{c}h_{a}{}^{e} + 18 \
h_{a}{}^{c} h^{ab} \partial_{b}h^{de} \partial_{c}h_{de} + 24 \
h_{a}{}^{c} h^{ab} h^{de} \partial_{c}\partial_{b}h_{de} - 24 \
h_{a}{}^{c} h^{ab} h^{de} \partial_{c}\partial_{e}h_{bd}  \nonumber\\
&&+ 36 h^{ab} \
h^{cd} \partial_{c}h_{a}{}^{e} \partial_{d}h_{be}
 - 24 h^{ab} h^{cd} \
\partial_{b}h_{a}{}^{e} \partial_{d}h_{ce} 
- 24 h_{a}{}^{c} h^{ab} \
h^{de} \partial_{e}\partial_{c}h_{bd} + 24 h_{a}{}^{c} h^{ab} h^{de} \
\partial_{e}\partial_{d}h_{bc} \nonumber\\
&&+ 24 h^{ab} h^{cd} \partial_{d}h_{ce} \
\partial^{e}h_{ab} - 6 h^{ab} h^{cd} \partial_{e}h_{cd} \
\partial^{e}h_{ab} 
- 24 h^{ab} h^{cd} \partial_{d}h_{be} \
\partial^{e}h_{ac} + 18 h^{ab} h^{cd} \partial_{e}h_{bd} \
\partial^{e}h_{ac}  \nonumber\\
&&- 24 h_{a}{}^{c} h^{ab} \partial_{c}h_{de} \
\partial^{e}h_{b}{}^{d} - 12 h_{a}{}^{c} h^{ab} \partial_{d}h_{ce} \
\partial^{e}h_{b}{}^{d} + 36 h_{a}{}^{c} h^{ab} \partial_{e}h_{cd} \
\partial^{e}h_{b}{}^{d},  \nonumber \\
%\ee
 %\vspace{-.8cm}
%\be
&& R_{\mu\nu}^{\left(1\right)}  =  \frac{1}{2}\left(-\partial_{\mu}\partial_{\nu}\underline{h}_{a}^{a}+\partial_{\mu}\partial^{a}\underline{h}_{a\nu}+\partial_{\nu}\partial^{a}\underline{h}_{a\mu}-\square\underline{h}_{\mu\nu}\right) \, , 
 \nonumber
 \\
&& R_{\mu\nu}^{\left(2\right)}  =  \frac{1}{2}\partial_{\mu}h^{ab}\partial_{\nu}h_{ab}+h^{ab}\left(\partial_{b}\partial_{a}h_{\mu\nu}+\partial_{\mu}\partial_{\nu}h_{ab}-\partial_{b}\partial_{\mu}h_{\nu a}-\partial_{b}\partial_{\nu}h_{\mu a}\right)+\partial^{b}h_{\mu}\phantom{}^{a}\left(\partial_{b}h_{\nu a}-\partial_{a}h_{\nu b}\right) \, , 
 \nonumber\\
&& R_{\mu\nu\varrho\sigma}^{(1)}  =  \frac{1}{2}\left(-\partial_{\mu}\partial_{\rho}h_{\nu\sigma}+\partial_{\mu}\partial_{\sigma}h_{\nu\rho}+\partial_{\nu}\partial_{\rho}h_{\mu\sigma}-\partial_{\nu}\partial_{\sigma}h_{\mu\rho}\right) \, ,
\ee
\be
&& \hspace{-1.4cm}
R_{\mu\nu\varrho\sigma}^{(2)}  =   - \tfrac{1}{2} \partial_{a}h_{\nu \sigma} \partial^{a}h_{\mu \rho} + \tfrac{1}{2} \partial_{a}h_{\nu \rho} \partial^{a}h_{\mu \sigma} + \tfrac{1}{2} \partial^{a}h_{\nu \sigma} \partial_{\mu}h_{\rho a} -  \tfrac{1}{2} \partial^{a}h_{\nu \rho} \partial_{\mu}h_{\sigma a} \nonumber\\
&& -  \tfrac{1}{2} \partial^{a}h_{\mu \sigma} \partial_{\nu}h_{\rho a} + \tfrac{1}{2} \partial_{\mu}h_{\sigma a} \partial_{\nu}h_{\rho}{}^{a}+ \tfrac{1}{2} \partial^{a}h_{\mu \rho} \partial_{\nu}h_{\sigma a} -  \tfrac{1}{2} \partial_{\mu}h_{\rho}{}^{a} \partial_{\nu}h_{\sigma a} \nonumber\\
&&+ \tfrac{1}{2} \partial_{a}h_{\nu \sigma} \partial_{\rho}h_{\mu}{}^{a} -  \tfrac{1}{2} \partial_{\nu}h_{\sigma a} \partial_{\rho}h_{\mu}{}^{a} -  \tfrac{1}{2} \partial^{a}h_{\mu \sigma} \partial_{\rho}h_{\nu a} + \tfrac{1}{2} \partial_{\mu}h_{\sigma a} \partial_{\rho}h_{\nu}{}^{a} \nonumber\\
&&-  \tfrac{1}{2} \partial_{a}h_{\nu \rho} \partial_{\sigma}h_{\mu}{}^{a} + \tfrac{1}{2} \partial_{\nu}h_{\rho a} \partial_{\sigma}h_{\mu}{}^{a} + \tfrac{1}{2} \partial_{\rho}h_{\nu a} \partial_{\sigma}h_{\mu}{}^{a} \nonumber \\
&&+ \tfrac{1}{2} \partial^{a}h_{\mu \rho} \partial_{\sigma}h_{\nu a} -  \tfrac{1}{2} \partial_{\rho}h_{\mu}{}^{a} \partial_{\sigma}h_{\nu a} -  \tfrac{1}{2} \partial_{\mu}h_{\rho a} \partial_{\sigma}h_{\nu}{}^{a} \, , \\
&& \hspace{-1.4cm}
R_{\mu\nu\varrho\sigma}^{(3)} =  - \tfrac{3}{2} h^{ab} \partial_{a}h_{\mu \sigma} \partial_{b}h_{\nu \rho} + \tfrac{3}{2} h^{ab} \partial_{a}h_{\mu \rho} \partial_{b}h_{\nu \sigma} -  \tfrac{3}{2} h^{ab} \partial_{a}h_{\nu \sigma} \partial_{\mu}h_{\rho b} + \tfrac{3}{2} h^{ab} \partial_{a}h_{\nu \rho} \partial_{\mu}h_{\sigma b} \nonumber\\
 &&-  \tfrac{3}{2} h^{ab} \partial_{\mu}h_{\sigma b} \partial_{\nu}h_{\rho a} + \tfrac{3}{2} h^{ab} \partial_{a}h_{\mu \sigma} \partial_{\nu}h_{\rho b} -  \tfrac{3}{2} h^{ab} \partial_{a}h_{\mu \rho} \partial_{\nu}h_{\sigma b} + \tfrac{3}{2} h^{ab} \partial_{\mu}h_{\rho a} \partial_{\nu}h_{\sigma b} \nonumber\\
 &&-  \tfrac{3}{2} h^{ab} \partial_{b}h_{\nu \sigma} \partial_{\rho}h_{\mu a} + \tfrac{3}{2} h^{ab} \partial_{\nu}h_{\sigma b} \partial_{\rho}h_{\mu a} -  \tfrac{3}{2} h^{ab} \partial_{\mu}h_{\sigma b} \partial_{\rho}h_{\nu a} + \tfrac{3}{2} h^{ab} \partial_{a}h_{\mu \sigma} \partial_{\rho}h_{\nu b} \nonumber\\
&&+ \tfrac{3}{2} h^{ab} \partial_{b}h_{\nu \rho} \partial_{\sigma}h_{\mu a} -  \tfrac{3}{2} h^{ab} \partial_{\nu}h_{\rho b} \partial_{\sigma}h_{\mu a} -  \tfrac{3}{2} h^{ab} \partial_{\rho}h_{\nu b} \partial_{\sigma}h_{\mu a} \nonumber \\
&&+ \tfrac{3}{2} h^{ab} \partial_{\mu}h_{\rho b} \partial_{\sigma}h_{\nu a} -  \tfrac{3}{2} h^{ab} \partial_{a}h_{\mu \rho} \partial_{\sigma}h_{\nu b} + \tfrac{3}{2} h^{ab} \partial_{\rho}h_{\mu a} \partial_{\sigma}h_{\nu b} .
\ee

\bibliographystyle{plain}
%\bibliography{MyAmplitudeBibliography}

\end{document}